\documentclass[reqno,11pt]{amsart}
\usepackage{geometry}
\usepackage{mathtools,amssymb,amsthm,mathrsfs,color,lineno,paralist,graphicx,float}
\usepackage[colorlinks,
linkcolor=blue,
anchorcolor=green,
citecolor=blue, 
]{hyperref}

\usepackage{enumitem}
\usepackage[T1]{fontenc}
\usepackage[utf8]{inputenc}
\usepackage{subfig} 
\usepackage[justification = centering, labelsep =period]{caption} 

\usepackage{calc}
\definecolor{bleu1}{RGB}{0,57,128}
\def\bleu1{\color{bleu1}}

\usepackage{etoolbox}
\patchcmd{\section}{\normalfont}{\normalfont \bleu1}{}{}
\patchcmd{\subsection}{\normalfont}{\normalfont \bleu1}{}{}
\patchcmd{\subsubsection}{\normalfont}{\normalfont \bleu1}{}{}
\renewcommand{\proofname}{\it \bleu1 Proof}

\newcommand{\SL}{\mathrm{SL}(2,\mathbb{R})}

\def\ti{\tilde}

\let\newpf\proof \let\proof\relax 
\newenvironment{pf}{\newpf[\proofname]}{\qed\endtrivlist}

\newcommand{\ba}{\overline{A}}

\newcommand{\bM}{\overline{M}}

\def\be{\begin{equation}}
\def\ee{\end{equation}}

\def\ba{{\begin{align}}}
\def\ea{{\end{align}}}

\def\bm{\begin{matrix}}
\def\em{\end{matrix}}

\def\u{{\mathbb U}}

\def\SL{{\mathrm{SL}}}
\def\PSL{{\mathrm{PSL}}}
\def\SO{{\mathrm{SO}}}

\def\0{{\mathbf 0}}

\newtheorem{Theorem}{Theorem}[section]
\newtheorem{Lemma}{Lemma}[section]
\newtheorem{Proposition}{Proposition}[section]
\newtheorem{Corollary}{Corollary}[section]
\newtheorem{Remark}{Remark}[section]

\numberwithin{equation}{section}

\def\ssm{\smallsetminus}
\def\tr{{\text{tr}}}
\renewcommand{\setminus}{\ssm}

\newcommand{\cl}{\mathbf{l}}
\def\cC{\mathcal{C}}
\newcommand{\inter}{\operatorname{int}}
\renewcommand{\mod}{\operatorname{mod}}

\newcommand{\id}{\operatorname{id}}

\newcommand{\A}{{\mathbb A}}
\newcommand{\C}{{\mathbb C}}

\newcommand{\Q}{{\mathbb Q}}
\newcommand{\R}{{\mathbb R}}
\newcommand{\T}{{\mathbb T}}
\newcommand{\V}{{\mathbb V}}

\newcommand{\Z}{{\mathbb Z}}

\def\B0{{\bold{0}}}

\def\bM{\bold{M}}

\def\bq{\bold{q}}

\def\bp{\bold{p}}

\def\bx{\bold{x}}
\def\by{\bold{y}}

\def\bw{\bold{w}}

\def\m{\bold{m}}

\def\P{\mathbb{P}}

\def\A{\bold{A}}

\def\U{\bold{U}}

\def\V{\bold{V}}

\catcode`\@=12

\def\Empty{}
\newcommand\oplabel[1]{
  \def\OpArg{#1} \ifx \OpArg\Empty {} \else
    \label{#1}
  \fi}

\newcommand{\comm}[1]{}
\newcommand{\comment}[1]{}

\begin{document}
\title[Dry Ten Martini Problem]{Dry Ten Martini Problem in the non-critical case}

\author{Artur Avila}

\address{Universit\"at Z\"urich, 
Institut f\"ur Mathematik\\
Winterthurerstrasse 190, 
CH-8057 Z\"urich, Switzerland  \& IMPA, Estrada Dona Castorina 110, Rio de Janeiro, Brazil\\}


\email{artur.avila@math.uzh.ch\\}

\author {Jiangong You}
\address{
Chern Institute of Mathematics and LPMC, Nankai University, Tianjin 300071, China} \email{jyou@nankai.edu.cn}

\author{Qi Zhou}
\address{
Chern Institute of Mathematics and LPMC, Nankai University, Tianjin 300071, China
}

 \email{qizhou@nankai.edu.cn}

%
%
%
%
\maketitle

\begin{abstract}
We solve the Dry Ten Martini Problem in the non-critical case, i.e.,   all possible spectral gaps are open for almost Mathieu operators with $\lambda\ne \pm 1$.

\end{abstract}

\tableofcontents

\setcounter{tocdepth}{1}


\section{Motivations and Main Results}

 This paper concerns the Almost Mathieu Operator (AMO):
\begin{equation}\label{schro}
(H_{\lambda,\alpha,x} u)_n= u_{n+1}+u_{n-1} +2\lambda \cos  2\pi  (
n\alpha + x) u_n,
\end{equation}
where $x\in \mathbb{T}$ is  the phase, $\alpha\in \R\backslash
\Q$ is the frequency and $\lambda\in \R \setminus \{0\}$
is  the coupling constant.
Without loss of generality, one can assume that
$\lambda>0$. The AMO was first introduced by Peierls \cite{Pe},
as a model for an electron on a 2D lattice, acted on by a homogeneous
magnetic field \cite{Ha,R}. This model
has been extensively studied not only because of  its importance
in  physics \cite{AOS,OA},
but also as a fascinating mathematical object.  Indeed, by varying the
parameters $\lambda, \alpha,x$, one sees surprising spectral
richness and it thus serves as a primary example for many spectral
phenomena \cite{L1}, which can be effectively analyzed due to some special
features and symmetries (such as Aubry duality) through a combination of
techniques stemming from different fields.

The spectrum of the self-adjoint operator
$H_{\lambda,\alpha,x}$ is a compact perfect set denoted by
$\Sigma_{\lambda,\alpha}\subset \R$, and it
is independent of $x$ for irrational 
$\alpha$.
   Each finite interval of  $\R\backslash\Sigma_{\lambda,\alpha}$ is called
an open gap, and it has an associated topologically defined
parameter in $[0,1]$.  This parameter
must satisfy the conclusions of the Gap Labelling Theorem (GLT), which
restricts it to a dense subset of $[0,1]$ labelled by integers.
However, it is in principle possible
for a label not to be
associated to an actual gap, and in this case one speaks of a closed gap.

In 1981,  during a talk at the   AMS annual meeting,
Mark Kac asked famously whether ``all gaps are there'' \cite{Kac,sim}, and
offered ten Martinis for the solution.  Barry Simon then popularized this
question in the form of two problems.  The first one,
called the Ten Martini Problem, asks whether the spectrum is a Cantor set
(this question actually dates back to Azbel \cite {Az}),
or equivalently, whether open gaps correspond to a dense set of parameters
in $[0,1]$.  The second more difficulty one, called the
Dry Ten Martini Problem, asks more precisely whether all gaps allowed by the
Gap Labelling Theorem are open.

%

Due to their central place in the spectral theory of
quasiperiodic Sch\"odinger operators, both problems
attracted lots of attention, with important progress being made by
Avila-Krikorian \cite {AK06}, Avila-Jitomirskaya \cite{AJ05,AJ08},
Bellissard-Simon \cite{BS},
Choi-Elliott-Yui \cite{CEY},
Helffer-Sj\"ostrand \cite{HS}, Last \cite{L},
Puig \cite{P},  Sinai \cite{Sin}.
For Cantor spectrum results concerning general quasiperiodic
Schr\"odinger operators, see also
Avila-Bochi-Damanik \cite{ABD} and Goldstein-Schlag \cite{GS11}. 

The Ten Martini Problem,
was completely solved in \cite{AJ05}, but
the Dry Ten Martini Problem remains open, and indeed up to this work it was
unknown whether its conclusion hold even for a single $\lambda$
(indeed all results so far
are dependent on certain arithmetic conditions on $\alpha$).

In this paper, we address \textit{completely} the noncritical case $\lambda \neq 1$.

\begin{Theorem}\label{dry}
$H_{\lambda,\alpha,x}$ has all spectral gaps open for all irrational
$\alpha$ and all $\lambda\neq 1$.
\end{Theorem}

Let us now describe the content of the GLT, so that
``all spectral gaps open'' becomes precisely defined.
We first recall that
the integrated density of states  $N_{\lambda,\alpha}(E)$ of
the operator $H=H_{\lambda,\alpha,x}$,  is defined as $$ N_{\lambda,\alpha}(E)=\int_{\R/\Z}
\mu_{\lambda,\alpha,x}(-\infty,E] dx, $$ where $
\mu_{\lambda,\alpha,x}$ is the spectral measure  associated with 
$H$ and $\delta_0$.  For each $E$,
the value of the
integrated density of states encodes some topological information about an
associated dynamical system, the fibered rotation number.

The integrated density
of states is continuous, strictly
monotonic restricted to the spectrum  and  locally constant in the complement of the spectrum.
The GLT \cite{BLT,DF,JM82} says that, in any fixed spectral gap,
there is a unique $k\in \Z \setminus \{0\}$ such that
$N_{\lambda,\alpha}$ is given by $k\alpha \,{\rm mod}\, \Z$ (note that
since $\alpha$ is irrational, different gaps must give rise to different
$k$).
The label of the spectral gap is just the integer $k$.

\begin{Remark}

The Dry Ten Martini Problem turns out to have relevance in physics as well. 
After Von Klitzing's discovery on quantum Hall
effect \cite{KDP}, Thouless and his coauthors \cite{TKNN}
gave a theoretic explanation by showing  that the Hall
conductance at the plateaus was related to a topological
invariant known as Chern number \footnote{See also the proof
from the methods of Non-Commutative Geometry \cite{BEB}.},
thus it is quantized (this is part of the theory for which Thouless
was awarded the 2016 Nobel Prize in Physics).

In physics language, the labels $k$ are identified with
Chern numbers and ``all
spectral gaps are open'' means that all topological phases
will show up for the quantum Hall system governed by the AMO. 

We note that some arguments of \cite {TKNN} assume that
all gaps of AMO are open when $\lambda$ varies.
Thus solving the Dry Ten Martini Problem for
$\lambda$ in an interval provides a mathematical justification for
the derivations in \cite{TKNN}. 

\end{Remark}

\subsection{Previous progress}\label{1.1}

Although in the statement of the Dry Ten Martini Problem,
the arithmetic property of
$\alpha$ is irrelevant, 
the methods developed to solve it
do  depend heavily on whether
$\alpha$ is
Diophantine or Liouvillean.  Here we use such concepts loosely to mean
``badly approximated by rational numbers'' and ``well approximated by
rational numbers'', with specific quantifications varying according to need.

The usual Diophantine condition regards polynomial
approximation, i.e., $\alpha$ is called Diophantine
if there exist $\gamma,\tau>0$ such that  $\|k\alpha\|_{\T} \geq
\frac{\gamma^{-1}}{|k|^{\tau}}, 0 \neq k \in \Z$.
The set of Diophantine $\alpha$ is denoted by $\mathrm {DC}$ (or
$\mathrm {DC}(\gamma,\tau)$ if the quantifiers are specified),
and $\alpha$ is said to be a
Liouvillean number if it is not Diophantine.

On the other hand, for problems related to the almost Mathieu operator, it
is often the exponential measures of approximation that play a more
important role.
More precisely, if   $\frac {p_n} {q_n}$ are 
the best approximations of $\alpha \in \R \setminus \Q$,  
$$ \beta=\beta(\alpha)=\limsup \frac {\ln q_{n+1}} {q_n},$$ 
can be used to measure how Liouvillean $\alpha$ is.
It is obvious that $\beta(\alpha)=0$ for (traditionally)
Diophantine $\alpha$.

In  1990, Choi, Elliott and  Yui \cite{CEY} developed a method which solved
the problem  for
some extremely Liouvillean $\alpha$. Optimizing the estimates in \cite{CEY},
Avila and Jitomirskaya \cite{AJ05} showed that if $\beta=\infty$ or if
$0<\beta<\infty$ and $e^{-\beta}<\lambda<e^{\beta},$ then
$H_{\lambda,\alpha,x}$ has all gaps open.
On the other hand, for Diophantine $\alpha$
there is a dynamical systems approach to
the Dry Ten Martini Problem based on the
{\it reducibility} of cocycles $(\alpha,A^{(\lambda,E)})$ associated to the
eigenvalue equation  $H_{\lambda,\alpha,x}u=Eu$ (reducibility essentially allows one
to analyze this equation as if the one-step transfer matrices were constant
along the lattice, that is, as easily as in the case of a
vanishing potential).\footnote{See Section \ref{preliminaries}
for the definitions.}
Here we say that two quasi-periodic cocycles $(\alpha, A^{i}),i=1,2$
are conjugated if there exists $B\in C^\omega(\T,\mathrm{PSL}(2,
\mathbb{R}))$ such that $$A^{1}(x)=B(x+\alpha)A^{2}
(x)B(x)^{-1},$$ and  $(\alpha, A)$  is \textit{reducible} if  it   is conjugated to a constant
cocycle.

In the Diophantine case (in the traditional sense, i.e., $\alpha \in \mathrm
{DC}$),  based on Eliasson's
reducibility result \cite{E92}, Puig \cite{P} solved the  Dry Ten
Martini Problem  in the perturbative regime, that is, for $\lambda$
small enough, with the smallness depends on the quantifiers $\gamma,\tau$ of
the Diophantine condition.
The strategy is
to prove that if $E$ is a possible gap boundary according to the GLT,
then $(\alpha,A^{(\lambda,E)})$ is analytically reducible  to a parabolic cocycle, and one use Moser-P\"oschel argument  \cite{MP84} to verify the openness of the gap.  Later on, 
Avila and Jitomirskaya developed a non-perturbative reducibility theory
which solved the problem for $\alpha\in
\mathrm {DC}$, and $\lambda\neq  1$ \cite{AJ08}.   It was later extended to $\beta(\alpha)=0$, and $\lambda\neq  1$ by Avila \cite{Aab}.

\subsection{Novelty, ideas of the proof}\label{1.2}

However, the above approaches from two sides can not solve the problem
completely for any fixed $\lambda$. In fact,  even after pushing such
approaches to their apparent technical limits, there would still remain an
arithmetically inaccessible range of parameters $|\ln \lambda| \in
[\beta,2\beta]$, see comments by Jitomirskaya
in \cite{J07}, and progress in \cite{LY}.

In this paper, we will introduce a new approach,  which works
for \textit{all} $\alpha$ and \textit{all} $0<\lambda<1$(this is called
the subcritical regime).  The case
$\lambda>1$ (the supercritical regime)
is an obvious consequence of the case $0<\lambda<1$ by Aubry
duality \cite{GJLS}.  
If $\lambda=1$, the associated operator is critical (this explains the title
of this paper).

We say  $(\alpha, A)$ is 
\textit{almost reducible} if the closure of its analytical conjugacy
class contains a constant.  It is known that  $(\alpha, A^{(\lambda,E)})$
are almost reducible for \textit{all} $\alpha$ and \textit{all}
$0<\lambda<1$ \cite{Aab,Aac,avila2}, however, reducibility can't be always
expectable (in fact, it is not reducible when $\alpha $ is very
Liouvillean \cite{AYZ}). 

The first key novelty is that we develop a
\textit{quantitative version of Aubry duality}, which reveals that
quantitative estimates of the conjugacy
imply quantitative estimates of the almost localized eigenfunctions of
the dual operators,  as a consequence, for  any  spectral gap boundary $E$, 
 $(\alpha, A^{(\lambda,E)})$
 can not be conjugated  to  a cocycle which is too close to the  identity.
The second novelty is that  we  develop  
 a  \textit{generalized Moser-P\"oschel argument} assuming only the
almost reducibility, and  prove the existence of some
$\tau' \in \R \setminus \{0\}$, such that the cocycle
$(\alpha,A^{(\lambda,E+\tau')})$ is
uniformly hyperbolic with $N_{\lambda,\alpha}(E+\tau')=N_{\lambda,\alpha}(E)$,
which implies the openness of the
corresponding gap. To validate the  \textit{quantitative Aubry dual}
and the  \textit{generalized Moser-P\"oschel argument},
we  develop \textit{quantitative
	almost reducibility}, which is based  on the global
almost reducibility theory, specifically the proof of the Almost Reducibility
Conjecture (ARC) developed  by Avila \cite{Aglobal,Aac,avila2},
and the  local almost reducibility theory developed by Hou-You \cite{HoY}.

We point out that even though the Dry Ten Martini problem was known already
in the Diophantine case, it is worthwhile to carry out the argument even in
that case since the new approach provides the first
quantitative estimates on the size of gaps for this case
(those are developed in those later works \cite{LYZZ}).

\subsubsection{Estimates on gaps of periodic approximants}

For the reader that is only concerned with verifying
the previously open Liouville
case, we provide a streamlined, almost self-contained version, which makes
use of periodic approximation to simplify some of the arguments.

If  the frequency is rational, say $\alpha=\frac{p}{q}$,
it is known that actually all gaps are non-collapsed, except when $q$ is
even in which case there is just one collapsed gap, the $q/2$-th, which
reduces to $\{0\}$.  Exponential lower bounds for the non-collapsed gaps
were given using
$C^*$-algebra techniques in \cite {CEY}, those were later
refined asymptotically in \cite {AJ05}.  Such results will not be
necessary in the following.

While there are indeed many exponentially (in $q$) small
gaps, we can show that there are
subexponential estimates for gaps with a given label
(c.f. Theorem \ref{subex}). 
To prove this, the following estimate plays a crucial role in our
argument.  It provides
specific exponential almost reducibility estimates for periodic cocycles in a particular
case (which for Schr\"odinger operators correspond to the presence of a
exponential small gap).

\begin{Theorem} \label {exp}

For every $0<\epsilon<\epsilon_0$, $c_0>0$ and $\delta>0$,
there exists $\delta_*>0$ and $q_*>0$ with the following property.
Let $p/q \in \Q$ and
$A \in C^\omega_{\epsilon_0}(\R/\Z,\SL(2,\R))$ be such that $q_*>q$ and
\be \label {sizek}
\sup_{0 \leq k \leq q} \frac {1} {q} \ln \|A_k\|_{\epsilon_0} \leq \delta_*,
\ee
where $A_k(x)=A(x+(k-1) p/q) \cdots A(x)$.  If $\|A_q(0)-I\| \leq e^{-c_0
q}$ with $I \in \{\id,-\id\}$ then there exists $B \in
C^\omega_\epsilon(\R/\Z,\SL(2,\R))$ and $R \in \SO(2,\R)$
such that $\|B\|_\epsilon \leq e^{\delta q}$, $R^q=I$, and $$\tilde
A(x)=B(x+p/q)A(x)B(x)^{-1}$$ satisfies
$\|\tilde A-R\|_0<e^{(-c_0+\delta) q}$ and $\|\tilde
A-R\|_\epsilon<e^{(-\gamma c_0+\delta) q}$ with $\gamma=1-\frac {\epsilon}
{\epsilon_0}$.  Moreover, $B$ may be chosen to be homotopic to a constant.

\end{Theorem}

General almost reducibility estimates for periodic
operators are given in
\cite {Aac} but provide only subexponential bounds.  Those can be refined to
exponential bounds through further KAM analysis (c.f. Section \ref{KAM-prop}).  We have
obtained a short proof for
this case of the ARC which gives directly the exponential estimates, this argument is given in Section \ref {exp-proof}.

\section{Preliminaries} \label{preliminaries}

 For a bounded
analytic (possibly matrix valued) function $F$ defined on $\T^d=\R^d/ \Z^d$ with an analytic extension to  $ \{ 
x \in \C^d/ \Z^d\  : \  | \Im x |< h \}$, we let
$
\|F\|_h=  \sup_{ | \Im x |< h } \| F(x)\|,$
(where $\|\cdot\|$ denotes absolute value or the usual matrix norm), and 
denote by $C^\omega_{h}(\T^d,*)$ the
set of all  those  $*$-valued bounded functions ($*$ will usually denote $\R$, $\mathrm{sl}(2,\R)$,
$\mathrm{PSL}(2,\R)$ and etc).  For any $F(x):= \sum_k   \widehat{F}(k) e^{ 2\pi i
\langle  k, x \rangle }\in C^\omega_{h}(\T^d,*)$, and any $L>0$,   $\mathcal{T}_L$ and $\mathcal{R}_L$  are used
to denote the truncation operators:
   $$ \mathcal {T}_L F= \sum_{|k|< L}
\widehat{F}(k)e^{2\pi  i
\langle  k, x \rangle }, \qquad \mathcal {R}_L F= \sum_{|k|\geq L}
\widehat{F}(k)e^{2\pi i
\langle  k, x \rangle }.$$
where for any $k=(k_1,k_2,\cdots, k_d)\in \Z^d$, we denote $|k|=|k_1|+ |k_2| +\cdots + |k_d|$.

\subsection{Cocycle,  Fibered rotation number}
A  cocycle $(\alpha, A)\in \R \times C^0(\T,
\mathrm{SL}(2,\R))$ is a linear skew product:
\begin{eqnarray*}\label{cocycle}
(\alpha,A):&\T \times \R^2 \to \T \times \R^2\\
\nonumber &(x,v) \mapsto (x+\alpha,A(x) \cdot v).
\end{eqnarray*}
 We define the products $(\alpha,A)^k=(k\alpha,A_k)$, where $A_0=\id,$
$$A_k(x)=A(x+(k-1)\alpha) \cdots
A(x), \qquad {\rm for } \ k \geq 1, $$ 
and define $A_{-k}=A_k(x-k\alpha)^{-1}$ for $k\ge 1$. If  $\alpha \in \R\backslash \Q$, we call $(\alpha,A)$ a  quasiperiodic cocycle, otherwise it is  a periodic cocycle.

We say that $(\alpha,A)$ is {\it uniformly hyperbolic} 
if $\|A_k(x)\|$ grows exponentially in $k$, uniformly in $x$.  In this case,
there exist unique
unstable and stable directions $u,s:\R/\Z \to \R/\Z$ which are
continuous, invariant in the sense that
$A(x) \cdot u(x)=u(x+\alpha)$, $A(x) \cdot s(x)=s(x+\alpha)$, and such that
$u(x) \neq s(x)$ everywhere (in particular they must be homotopic).  They
are distinguished so that $A_k$ contracts exponentially
vectors in $s$ and expands exponentially vectors in $u$.   By the conefield criterion, the set of uniformly hyperbolic cocycles is  an open set  in $\R \times C^0(\R/\Z , \mathrm{SL}(2,\R))$.

Assume now that $A:\T \to \mathrm{SL}(2,\R)$ is homotopic to the identity,
then  the continuous map
\begin{equation*}
\begin{array}{rrcl}
F: & \T \times \{v: \|v\|=1\}  & \longrightarrow     &  \T \times  \{v: \|v\|=1\}  \\
      & (x,v)    & \mapsto & \left(x+\alpha,
      \frac{A(x)v}{\| A(x) v \|}  \right)
\end{array}
\end{equation*}
is also homotopic to the identity. Therefore, it admits a continuous
lift $\tilde{F}: \T \times \R  \to  \T \times \R$ of the form $
\tilde{F}(x,t)= \left(x + \alpha, t+ f(x,t) \right)
$ with 
$$
f(x +1,t +1)= f(x,t) \quad\mbox{and} \quad p\left(t +
f(x,t)\right)= \frac{A(x)p(t)}{\| A(x) p(t) \|}
$$
for all $t \in \R$ and $x \in \T$,  where  $p:\R \to \T $ is defined as  $p(t)= e^{ 2\pi  i t}$.   The map $f$ is independent
of the choice of $\tilde{F}$ up to  a constant $k \in \Z$. Since $x \mapsto x + \alpha$ is uniquely ergodic on
$\T$,  the limit
\[
\lim_{N \to \infty} \frac{1}{ N} \sum_{n=0}^{N-1}
f\left(\tilde{F}^n(x,t)\right)
\]
exists modulus $\Z$ and  is independent of $(x,t)$ \cite{H,JM82}. 
This
limit is called the fibered rotation number of $(\alpha, A)$  and
 will be denoted by $\rho_f(\alpha, A)$. 

Any $A\in
C^0(\T, \PSL(2,\R))$ is homotopic to $x \mapsto R_{nx}$ for
some $n\in\Z$, where $$R_{\phi}=\left (\bm \cos 2\pi  \phi&-\sin  2\pi  \phi\\
\sin 2\pi  \phi &\cos 2\pi    \phi \em \right ).$$
We call $n$ the  topological degree of $A$, and denote it by $\deg A
$.  If two cocycles $(\alpha,A_1)$  and  $(\alpha,A_2)$ are conjugated, i.e.,  $
B(x+\alpha)^{-1}A_1(x)B(x)=A_2(x)$ for some $B
\in C^\omega(\T,$ $\mathrm{PSL}(2,\R))$ with  $\deg B=n \in\Z$, then
\begin{equation}\label{rot-conj}
\rho_f(\alpha, A_1)= \rho_f(\alpha, A_2) +\frac{1}{2}  n \alpha \quad \mod \Z.
\end{equation}
From the definitions, we also  have
\begin{equation}\label{rot-diff} |\rho_f(\alpha,A)-\phi|<C\|A-R_{\phi}\|_0.
\end{equation}

\subsection{Schr\"odinger operator, Integrated density of states}

Given a potential $v \in C(\R/\Z,\R)$ and $\alpha \in \R$ we can consider,
for each $x \in \R/\Z$ the Schr\"odinger operator
$$(H_vu)_n=u_{n+1}+u_{n-1}+v(x+n \alpha) u_n.$$
Note that a sequence $(u_n)_{n \in \Z}$ is a formal solution of the
eigenvalue equation $H_{v} u=Eu$ if and only if
it satisfies $$\begin{pmatrix}
u_{n+1}\\u_n\end{pmatrix}=A^{(E)} (x+n\alpha) \cdot
\begin{pmatrix} u_n\\u_{n-1} \end{pmatrix},$$ where
$A^{(E)}=\begin{pmatrix} E-v&-1\\1&0 \end{pmatrix}$, and we call $(\alpha,A^{(E)})$ a Schr\"odinger cocycle.  Thus corresponding to the AMO, this naturally defined cocycle is also called almost Mathieu cocycle, where we denote  
\be \label {alambdae}
A^{(\lambda,E)}(x)=\begin{pmatrix} E-2 \lambda \cos 2 \pi x & -1
\\ 1 & 0 \end{pmatrix}.
\ee

Denotes the integrated density of
states  of the Sch\"odinger operator  $H$  by  $N$,
it  is known that the support of the probability measure $dN$ is the spectrum
of $H$, it relates to the fibered rotation number  in the following way
\begin{equation}\label{rot-ids}
N(E)=1-2 \rho_f(\alpha,A^{(E)}).
\end{equation}
 For $\alpha$ irrational, by a result of Johnson \cite{J},  the spectrum consists of the energies $E$ for which the Schr\"odinger cocycle $(\alpha,A^{(E)})$ is not uniformly hyperbolic:
 \begin{eqnarray}\label{uhequi}
\R \backslash \Sigma=\{E\in \R\quad | \quad  (\alpha, A^{(E)}) \quad \text{is uniformly hyperbolic} \}
\end{eqnarray}

%
%

\subsection{Global theory of one frequency quasi-periodic  $SL(2,\R)$ cocycle}

For any $\alpha \in \R\backslash \Q$,
the Lyapunov exponent of $(\alpha,A)$ is
defined as
$$ L(\alpha, A)=\lim_{n\rightarrow \infty} \frac {1} {n}
\int \ln \|A_n(x)\| dx.
$$
Suppose that  $A\in$ $C_{\delta}^\omega(\R/\Z,\SL(2,\R))$ admits a
holomorphic extension to $|\Im x|<\delta$, then for
$|\epsilon|<\delta$ we can define $A_\epsilon \in
C^\omega(\R/\Z,\SL(2,\C))$ by $A_\epsilon(x)=A(x+i \epsilon)$.
 The cocycles which are not uniformly hyperbolic are classified 
 into three regimes: subcritical, critical, and supercritical. In
 particular, $(\alpha, A)$ is said to be
 subcritical, if there exists $\varepsilon_0>0,$ such that
 $L(\alpha,A_{\varepsilon})=0$ for  all $|\varepsilon|<\varepsilon_0.$

 Concerning the Almost Mathieu cocycle, we have the following: 
 
 \begin{Theorem}\cite{Aglobal}\label{bj-formula}
If $\alpha \in \R \backslash \Q$, $\lambda<1$, $E \in \R$, then for
$\epsilon \geq 0$,
$$L(\alpha,(A^{(\lambda,E)})_\epsilon)=\max
\{L(\alpha,A^{(\lambda,E)}),( 2\pi \epsilon+\ln \lambda)\}.$$
It follows that 
\begin{equation}\label{bj-f-1}L(\alpha,A^{(\lambda,E)})=\max
\{0,\ln |\lambda|\}, \quad \forall E \in \Sigma_{\lambda,\alpha}. \end{equation}
\end{Theorem}
\begin{Remark}
 \eqref{bj-f-1} was first proved by Bourgain-Jitomirskaya \cite{BJ}. 
\end{Remark}

\section{Quantitative Almost Reducibility}

The Almost Reducibility Conjecture (ARC) is a result of global type
stating that  a cocycle $(\alpha, A)$  is almost reducible if it is
subcritical \cite{Aglobal}.
ARC was announced in \cite{Aglobal}, and proved in \cite{Aac,avila2}.

In this section we provide quantitative
estimates for almost reducibility of cocycles associated to gap boundaries
by applying local techniques after a global to local reduction based on the
original ARC estimates.

In the Liouvillean case  $\beta(\alpha)>0$, we start from the following:

\begin{Theorem}[Avila \cite{Aac}]\label{arc}
Let $\alpha\in\R\backslash \Q$ with $\beta(\alpha)>0$, and suppose that  $(\alpha,
A)$ is subcritical  in  $| \Im x |< h $. Then for
sufficiently small $\delta^{'}>0$,  there exist  $C=C(h)>0$, $h_*=h_*(C,\delta^{'},h)>0$,
a subsequence of $q_n$ with $q_{n+1}> e^{(\beta-o(1))q_n}$, $W_n(x)\in C_{h_*}^\omega(\T,\mathrm{PSL}(2,\R))$ with $\deg W_n=0$
 and $R_{\varphi_n}\in
SO(2,\R)$ such that \begin{eqnarray}
W_n(x+\alpha)^{-1}A(x)W_n(x)=
R_{\varphi_n}+G_n(x)
\end{eqnarray}
where  $\|W_n\|_{h_*}\leq
e^{Cq_{n}\delta^{'}}$, $\|G_n\|_{h_*}\leq e^{-{q_{n}\delta^{'}}}$.
\end{Theorem}

The estimates in Theorem \ref{arc} are not quite enough for our application
because the size of the conjugacy necessary to get $\epsilon$-close to a
constant is too large, of order $\epsilon^{-C}$. 
What we call \textit{quantitative almost reducibility} is the estimate
provided in
Theorem \ref{prop-main} below, where  $\|W_n\|_h\cdot \|G_n\|_h$ tends
to zero at some rate\footnote{We remark that to get H\"older continuity of
the rotation number,  a nice estimate on
$\|W_n\|_{C^0}\cdot \|G_n\|_{C^0}$ ($C^0$ norm) is
enough \cite{AJ08} ,  but here we do need nice  estimates on the
analytic norm.} when the rotation number is
rational w.r.t $\alpha$, i.e.  $2\rho_f(\alpha,A)=k_0\alpha\  mod\  \Z$. 
In order to do this, the constant cocycle we aim to approach can not be
taken as a rotation, but in triangular form (shrinking the diagonal element
can of course be done, but is costly).

\def\cL{\mathfrak{L}}
We first introduce some notations:
\begin{eqnarray*}
\varepsilon_n&=&2e^{\frac{2\pi h_*}{1+\alpha}}
e^{-q_{n}\delta^{'}/2},\\
\delta_n&=& h_*/2^n,\\
h_n&=&h_*/(1+\alpha)-\delta_n/6,
\end{eqnarray*}
and also
 $$\cL=
\left(
\begin{array}{ccc}
0& 1\\
0 &  0
 \end{array}\right).$$
Then we have the following:

\begin{Theorem}\label{prop-main}
Let $\alpha\in\R\backslash \Q$ with $\beta(\alpha)>0$, 
$k_0\in\Z$ and  suppose that  $(\alpha, A)$ is subcritical and
$\cl=2\rho_f(\alpha,A)-k_0\alpha \in \Z$. Then there exist a subsequence of
$q_n$ with $q_{n+1}> e^{(\beta-o(1))q_n}$, $B_n(x)\in C_{h_n}^\omega(\T, $
$\mathrm{PSL}(2,\R))$ with  $\deg B_n=k_0$ and $d_n
\in \R$ such that
\begin{eqnarray}
(-1)^\cl B_n(x+\alpha)^{-1}A(x)B_n(x)= \id+d_n \cL+F_n(x)
\end{eqnarray}
with estimates   $\|B_n\|_{h_n}\leq
e^{2q_{n+1}\varepsilon_n^{\frac{1}{4}}}$,
$|d_n|\leq e^{-q_{n+1}\varepsilon_n^{\frac{1}{4}}}$ and $\|F_n\|_{h_n}\leq
e^{-q_{n+1}\delta_n}$.
\end{Theorem}

\begin{Remark}
Note the selection $q_{n+1}> e^{(\beta-o(1))q_n}$ implies that
$q_{n+1}\gg \varepsilon_n^{-\frac{1}{4}}$, then
$|d_n|$ is actually small. Also $\delta_n \gg \varepsilon_n^{\frac{1}{4}}$, then the size of the perturbation $F_n(\cdot)$ is much smaller than $|d_n|$.

\end{Remark}

For the case $\beta(\alpha)=0$, a qualitative result is enough:

\begin{Theorem}\label{prop-main-dio}\cite{Aab,avila2}
Let $\alpha\in\R\backslash \Q$ with $\beta(\alpha)=0$, 
$k_0\in\Z$ and  suppose that  $(\alpha, A)$ is subcritical and
$\cl=2\rho_f(\alpha,A)-k_0\alpha \in \Z$. Then $(\alpha,A)$ is reducible, i.e.
there exists $B(x)\in C^\omega(\T,\mathrm{PSL}(2,\R))$, $d \in \R$ such that 
\begin{eqnarray}
(-1)^\cl B(x+\alpha)^{-1}A(x)B(x)=\id+d \cL.
\end{eqnarray}

\end{Theorem}

\begin{Remark}

For the case of interest here this result
follows directly from the Theorem 3.8 of \cite {Aab} which applies to
all subcritical Almost Mathieu cocycles.  For general
cocycles, the results of
\cite {Aab} only covers the case close to constant, so
a global to local reduction must be done
preliminarily, using the ARC \cite {avila2}.

\end{Remark}

\begin{Remark}

While we have not stated it in the above formulation,
both \cite {Aab} and \cite
{avila2} provide quantitative estimates on the size of conjugacies, so can
be used for bounds on gap size, see \cite {LYZZ}.

\end{Remark}

The remaining of this section is dedicated to the proof of
Theorem \ref {prop-main} :

 \subsection{Poincar\'e cocycle and local embedding theorem} 
 
To prove Theorem \ref{prop-main}, we will first embed the cocycle into a
quasiperiodic linear
 system, perform the KAM scheme for the embedded linear system,
then get desired estimates for the cocycle.  Here,  we 
 consider the analytic quasiperiodic linear
 system 
 \begin{eqnarray}\label{qp-0}
\left\{ \begin{array}{l}\dot{\bx}=A(\varphi)\bx \\
\dot{\varphi}=\omega=(1,\alpha)
\end{array} \right.
\end{eqnarray}
 where  $A\in
C^\omega_{h}(\T^2,sl(2,\R))$.  Denote $\Phi^t(\varphi)$  the corresponding fundamental solution matrix
of $(\ref{qp-0})$, and  denote $\Phi^1(\varphi)=\Phi^1(0,x)= \tilde A(x)$, then we call $(\alpha, \tilde A)$ the Poincar\'e cocycle associated with \eqref{qp-0}. 
Recall that \eqref{qp-0} is analytically conjugated to 
 \begin{eqnarray}\label{qp-00}
\left\{ \begin{array}{l}\dot{\by}=A'(\varphi)\by \\
\dot{\varphi}=\omega=(1,\alpha)
\end{array} \right.
\end{eqnarray}
if there exists $B\in
C^\omega(\T^2, \mathrm{PSL}(2,\R))$, such that \eqref{qp-0} can be transformed by the change of variables $\bx\rightarrow B(\varphi) \by$ into \eqref{qp-00}. Then similar as in the cocycle case, \eqref{qp-0} is {\it almost reducible}, if the closure of its analytical conjugated class contains a constant.   One of the basic relation is the following:

\begin{Theorem}\label{equi-ar}\cite{YZ}
 An analytic  quasi-periodic linear system \eqref{qp-0} is almost reducible if and only if its corresponding Poincar\'e cocycle
$(\alpha, \tilde A)$ is almost reducible.
\end{Theorem}

Conversely,  if  $(\alpha, \tilde A)$ is almost reducible, particularly  if $\tilde A(x)$ is close to constant, then  $(\alpha, \tilde A)$ can be embedded into an analytic quasi-periodic linear
 system.

\begin{Theorem} \cite{YZ}\label{localemb}
Suppose that $\alpha\in\R\backslash \Q$, $h>0,$ $G\in
C^\omega_{h}(\T,\mathrm{sl}(2,\R))$, $A\in \mathrm{sl}(2,\R)$ being constant. Then
there exist $\epsilon=\epsilon(A,h,|\alpha|)>0,$
$\widetilde{\eta}=\widetilde{\eta}(A,h,|\alpha|)>0,$
$\widetilde{A}\in \mathrm{sl}(2,\R)$ and $F\in
C^\omega_{h/1+|\alpha|}(\T^2,\mathrm{sl}(2,\R))$ such that the cocycle
$(\alpha,e^A e^{G(\cdot)})$ is the Poincar\'e map of
\begin{eqnarray}
\left\{ \begin{array}{l}\dot{\bx}=(\widetilde{A}+F(\varphi))\bx \\
\dot{\varphi}=\omega=(1,\alpha)
\end{array} \right.
\end{eqnarray}
provided that $\|G\|_{h}<\epsilon.$ Moreover,  the following is true
\begin{enumerate}
\item If $A$ is in
the real normal forms $\left(
\begin{array}{ccc}
 \lambda &  0\cr
 0 &  -\lambda\end{array} \right)$ or
  $\left(
\begin{array}{ccc}
0  &  \rho\cr
 -\rho &  0\end{array} \right),$ then $\widetilde{A}=A,$ $\|F\|_{\frac{h}{1+|\alpha|}}\leq 2\widetilde{\eta} \|G\|_{h}$. In this case,  $\epsilon$ and  $\widetilde{\eta}$ can be taken as $\epsilon=\frac{2}{\pi^2}e^{-\frac{4\pi
h(2|\rho|+1)}{1+|\alpha|}}$, $\widetilde{\eta}= e^{\frac{2\pi
h(2|\rho|+1)}{1+|\alpha|}}$.
 \item If
 $A$ is in the real normal form $\left(
\begin{array}{ccc}
 0 &  1\\
 0 &  0
 \end{array}\right),$ then $\widetilde{A}=\left(
\begin{array}{ccc}
 0 &  0\\
 0 &  0
 \end{array}\right)$ and $\|F\|_{\frac{h}{1+|\alpha|}}\leq 2\widetilde{\eta} \|G\|_{h}^{\frac 12}.$
\end{enumerate}

\end{Theorem}

\subsection{Proof of Theorem \ref{prop-main}}\label{KAM-prop}

From now on, $k_0\in\Z$ is arbitrary but fixed.  Let $q_n$ be the selected
sequence as in Theorem \ref{arc}.
Furthermore we  assume that 
\begin{equation}\label{sel-k}
\quad q_n>   2(|k_0|+1)h_*/\delta^{'}.\end{equation}
 
\begin{Lemma}\label{local-reduction}
Let $\alpha\in\R\backslash \Q$ with $\beta(\alpha)>0$,
$k_0\in\Z,$ and  suppose that  $(\alpha, A)$ is subcritical with
$\cl=2\rho_f(\alpha,A)-k_0\alpha \in \Z$. Then
 there exists  $\overline{W}_n(x)\in C_{h_*}^\omega(\T, \mathrm{PSL}(2,\R))$ with  $\deg \overline{W}_n=k_0 $, such that
\begin{eqnarray}
(-1)^{\cl} \overline{W}_n(x+\alpha)^{-1}A(x)\overline{W}_n(x)=e^{\widetilde{G}_n(x)}
\end{eqnarray}
with estimate $\|\overline{W}_n\|_{h_*}\leq e^{Cq_{n}\delta^{'}}$ and $\|\widetilde{G}_n\|_{h_*}\leq
2e^{-q_{n}\delta^{'}/2}$.
\end{Lemma}

\begin{pf}
For definiteness, we consider the case $(-1)^\cl=1$, the other being analogous.
By the assumption and Theorem \ref{arc}, $(\alpha, A)$ is conjugated to
$(\alpha,R_{\varphi_n}+G_n(x))$ by $(0,W_n(x))$. Since $\deg
W_n=0$,  then by $(\ref{rot-conj})$, we have $2\rho_f(\alpha,
R_{\varphi_n}+G_n(x))=k_0\alpha \quad \text{mod}\Z.$ Let $\overline{W}_n(x)= W_n(x)
R_{k_0x/2}$.  Then
\begin{eqnarray}\label{con-1} \overline{W}_n(x+\alpha)^{-1}A(x)\overline{W}_n(x)=
R_{\varphi_n-k_0\alpha/2}+\overline{G}_n(x)
\end{eqnarray}
with 
$2\rho_f(\alpha, R_{\varphi_n-k_0\alpha/2}+\overline{G}_n(x))=0 \ \text{mod} \ \Z.$
Moreover, by our selection \eqref{sel-k}, we have estimates:
\begin{eqnarray*} \|\overline{W}_n\|_{h_*}\leq e^{Cq_{n}\delta^{'}}, \quad
\|\overline{G}_n\|_{h_*}\leq e^{-q_{n}\delta^{'}/2}.
\end{eqnarray*}

By $(\ref{rot-diff})$, it follows that $$|
\varphi_n-\frac{k_0\alpha}{2}| \leq \|\overline{G}_n\|_{h_*}\leq
e^{-q_{n}\delta^{'}/2},$$ which means that $(\alpha,R_{\varphi_n-k_0\alpha/2} )$ can seen as  a $e^{-q_{n}\delta^{'}/2}$-perturbation of the  cocycle $(\alpha,\id)$.  Thus we can  rewrite $(\ref{con-1})$ as
\begin{eqnarray}\label{con-2} \overline{W}_n(x+\alpha)^{-1}A(x)\overline{W}_n(x)=R_{\varphi_n-k_0\alpha/2}+\overline{G}_n(x)=
e^{\widetilde{G}_n(x)}
\end{eqnarray}
with estimate $\|\widetilde{G}_n\|_{h_*}\leq
2e^{-q_{n}\delta^{'}/2}$.

\end{pf}

Next we  embed the local
cocycle $(\alpha,e^{\widetilde{G}_n(x)})$ into a quasi-periodic
linear system \cite{YZ}, and then use Hou-You's method \cite{HoY} to obtain finer estimates for the embedded linear system. Consequently we obtain  finer estimates for initial  cocycle $(\alpha,e^{\widetilde{G}_n(x)})$. We will explain in Remark \ref{reason} why  this indirect approach is needed.  

By the definition of $q_n$, we have $$ 2e^{-q_{n}\delta^{'}/2}<
\frac{2}{\pi^2}e^{-\frac{4\pi h_*}{1+\alpha}}.$$ Then by Theorem
\ref{localemb}, the cocycle $(\alpha,
e^{\widetilde{G}_n(x)})$ can be embedded into an analytical quasi-periodic linear system
\begin{eqnarray}\label{al-ref1}
\left\{ \begin{array}{l}\dot{\bx}=\widehat{F}_n(\varphi)\bx \\
\dot{\varphi}=\omega=(1,\alpha)
\end{array} \right.
\end{eqnarray}
where $\varphi\in\T^2,$ $\widehat{F}_n\in
C^\omega_{\frac{h_*}{1+\alpha} }(\T^2, \mathrm{sl}(2,\R))$ with estimate
$$\|\widehat{F}_n\|_{\frac{h_*}{1+\alpha}}\leq 2e^{\frac{2\pi
h_*}{1+\alpha}} e^{-q_{n}\delta^{'}/2}=  \varepsilon_n  .$$  
Note that the fibered rotation number of the system \eqref{al-ref1}  satisfies 
\begin{equation}\label{rot2} 2\rho_f(\omega,\widehat{F}_n(\varphi))=0 \mod \Z \end{equation} since by Lemma \ref{local-reduction} and \eqref{rot-conj}, we have $2\rho_f(\alpha,e^{\widetilde{G}_n(x)})=0 \mod \Z$.
 In order to make the cocycle $(\alpha,
e^{\widetilde{G}_n(x)})$ closer to constants, by
Theorem \ref{equi-ar}, we only need
to conjugate the quasi-periodic linear  system
$(\ref{al-ref1})$ to a system closer to constants. This can be achieved by  Hou-You's method \cite{HoY}  for the almost
reducibility of local quasi-periodic systems. In fact, only  one cycle of  Hou-You's iteration is needed  for this purpose.

\begin{Proposition}\label{prop-hy}
There exist   $W_n^{'}\in
C_{h_n}^\omega(\T^2, \mathrm{PSL}(2,\R))$ with $\|W_n^{'}\|_{h_n}\leq
e^{2q_{n+1}\varepsilon_n^{\frac{1}{4}}}$, $\deg W_n^{'}=0,$ such
that $W_n^{'}(\varphi)$ further conjugates $(\ref{al-ref1})$ to
\begin{eqnarray}\label{al-ref2}
\left\{ \begin{array}{l}\dot{ \bx}=(d_n \cL+F_{n}^{'}(\varphi))\bx \\
\dot{\varphi}=\omega=(1,\alpha)
\end{array} \right.
\end{eqnarray}
 with estimate $|d_n|\leq e^{-q_{n+1}\varepsilon_n^{\frac{1}{4}}}$
 and $\|F_{n}^{'}\|_{
h_n}\leq e^{-q_{n+1}\delta_n}$.
\end{Proposition}

The estimates in Proposition \ref{prop-hy} is more precise than that in  \cite{HoY}, its proof is a modified version of their proof. In the following, we will first finish the proof of Theorem \ref{prop-main},
while postpone the proof of Proposition \ref{prop-hy} a bit later. \\

Since $(\alpha, e^{\widetilde{G}_n(x)})$ is the Poincar\'e cocycle
of the quasi-periodic linear system $(\ref{al-ref1})$, by Proposition \ref{prop-hy} and Theorem
\ref{equi-ar}, we obtain  the following:

\begin{Corollary}\label{dis-local}
There exists
 $B_n^{'}(x)\in
C_{h_n}^\omega(\T,\mathrm{PSL}(2,\R))$ with $\|B_n^{'}\|_{h_n}\leq
e^{2q_{n+1}\varepsilon_n^{\frac{1}{4}}}$, $\deg B_n^{'}=0,$ such that
\begin{eqnarray}\label{conj5}
B_n^{'}(x+\alpha)^{-1}e^{\widetilde{G}_n(x)}B_n^{'}(x)=\id+d_n \cL+F_n(x)
\end{eqnarray}
where   $|d_n|\leq e^{-q_{n+1}\varepsilon_n^{\frac{1}{4}}}$. Moreover,  $\|F_n\|_{h_n}\leq
e^{-q_{n+1}\delta_n}$.
\end{Corollary}

Once we have this, let $B(x)= \overline{W}_n(x)  B_n^{'}(x)$.  Then Theorem \ref{prop-main} follows from Corollary \ref{dis-local} directly.  \qed

\smallskip
\textbf{Proof of Corollary \ref{dis-local}:}
By Proposition \ref{prop-hy} and Theorem \ref{equi-ar},  $(\alpha, e^{\widetilde{G}_n(x)})$ is almost reducible. Moreover, quantitative estimates are inherited from the quasi-periodic linear system, this is essentially contained  in Lemma $5.1$ of \cite{YZ}. We sketch the proof here for completeness. 

Let $\Phi^t(\varphi)$ and $\ti \Phi^t(\varphi)$ be the corresponding fundamental solution matrix
of $(\ref{al-ref1})$ and  $(\ref{al-ref2})$,  one has
\begin{equation}\label{al-ref3-1}\ti \Phi^t(\varphi)=e^{d_n \cL t}\Big(\id +
\int_0^t e^{-d_n \cL s }F'_n(\varphi+s\omega) \ti
\Phi^s(\varphi)ds \Big).\end{equation}
Denote by $G^t(\varphi)=e^{-d_n \cL t} \ti \Phi^t(\varphi)$, then
$$
G^t(\varphi)=\id +\int_0^t e^{-d_n \cL s } F_n'(\varphi+\omega
s)e^{d_n \cL s }G^s(\varphi)ds,
$$
let $g(t)=\|G^t(\varphi)\|_{h_n}$, then
$$
g(t) \leq 1 + \int_0^t \|F'_n\|_{h_n} (1+ |d_n| s)^2 g(s) ds.
$$
By Gronwall's inequality,  for $0\leq t\leq 1$, we have  $g(t)\leq e^{2 \|F'_n\|_{h_n} t}.$

In  $(\ref{al-ref3-1})$, let $t=1$, then we have $$\ti
\Phi^1(0,x)=(\id+d_n \cL)(\id+\widetilde{F}_n(x)),$$ with the estimate
$$\|\widetilde{F}_n\|_{h_n}\leq \int_0^1\|F'_n\|_{h_n} (1+ |d_n| s)^2  g(s) ds\leq
2\|F_n\|_{h_n}.$$ Since $W_n^{'}(\varphi)$  conjugates $(\ref{al-ref1})$ to
$(\ref{al-ref2}):$
\begin{equation}\label{flow-po}\Phi^t(0,x)=W_n^{'}(t,x+t\alpha)\ti \Phi^t(0,x)
W_n^{'}(0,x)^{-1}.\end{equation} Let $t=1$, $B_n^{'} (x)= W_n^{'}(0,x) $, then \eqref{conj5} follows from \eqref{flow-po}, and the rest estimates follow from Proposition \ref{prop-hy}. 
\qed

\subsection{Proof of Proposition \ref{prop-hy}:}
For any given $h>0$, $\eta>0$, $\omega=(1,\alpha)\in \mathbb{R}^2$
and $A\in \mathrm{sl}(2,\mathbb{R})$, we decompose
$C_{h}^\omega(\T^2,\mathrm{sl}(2,\R))=\mathfrak{B}^{(nre)}_h\oplus
\mathfrak{B}^{(re)}_h$ (The decomposition depends on $A$, $\omega$,
$\eta$) in
  such a way that for any $Y\in
\mathfrak{B}_h^{(nre)}$
\begin{eqnarray*}
\partial_\omega Y, [A,Y]\in
\mathfrak{B}_h^{(nre)},\quad |\partial_{\omega}Y-[A,Y]|_h\geq \eta
|Y|_h. \end{eqnarray*}

In the present case, $A=0$  and $ \eta=
\varepsilon_n^{\frac{1}{4}}$.  We first have the following:

\begin{Lemma}
There exists  $B_1\in
C^\omega_{\frac{h_*}{1+\alpha}}(\T^2,\SL(2,\R))$ with $ \|B_1
-\id\|_{\frac{h_*}{1+\alpha}}\leq 2\varepsilon_n^{\frac{1}{4}},$
such that $B_1(\varphi)$ conjugates the system (\ref{al-ref1}) to
\begin{eqnarray}\label{al-ref3}
\left\{ \begin{array}{l}\dot{\bx}=F_1^{(re)}(\varphi)\bx \\
\dot{\varphi}=\omega=(1,\alpha)
\end{array} \right.
\end{eqnarray}
where $F_1^{(re)}(\varphi)\in \mathfrak{B}^{(re)}_{ \frac{h_*}{1+\alpha}}$
satisfying $  \|F_1^{(re)}(\varphi)\|_{\frac{h_*}{1+\alpha}}\leq
2\varepsilon_n.$ Moreover, we have
\begin{equation}\label{rot1}
2 \rho_f(\omega,F_1^{(re)}(\varphi))=0 \mod \Z
\end{equation}
and $F_1^{(re)}(\varphi)$ has the form
\begin{eqnarray}\label{1111}
\mathcal{T}_{\frac{q_{n+1}}{6}}F_1^{(re)}(\varphi)=
\sum_{k=l(-p_n,q_n), |k|< \frac{q_{n+1}}{6}}F_1^{(re)} (k)
e^{ 2\pi i\langle k,\varphi \rangle}.
\end{eqnarray}
\end{Lemma}

\begin{pf}
See Lemma $5.1$ of \cite{HoY}.  Note that $\deg B_1=0$ since $ \|B_1 -\id\|_{\frac{h_*}{1+\alpha}}\leq
2\varepsilon_n^{\frac{1}{4}}$, consequently \eqref{rot1} follows from \eqref{rot-conj} and \eqref{rot2}.
\end{pf}

\begin{Remark}
This step can be applied to discrete  cocycles almost without change, which will be used once again in the final proof (Section \ref{section5}).
\end{Remark}

\begin{Remark} Due to the special form of  $\mathcal{T}_{\frac{q_{n+1}}{6}}F_1^{(re)}(\varphi)$,  the rotation step as in Lemma $5.2$ of
\cite{HoY} is not necessary. So one doesn't need to shrink the
analytical radius greatly. 
\end{Remark}

The special form (\ref{1111}) of
$\mathcal{T}_{\frac{q_{n+1}}{6}}F_1^{(re)}(\varphi)$ implies that
the system
\begin{eqnarray}\label{al-ref3'}
\left\{ \begin{array}{l}\dot{\bx}=\mathcal{T}_{\frac{q_{n+1}}{6}}F_1^{(re)}(\varphi)\bx \\
\dot{\varphi}=\omega=(1,\alpha)
\end{array} \right.
\end{eqnarray}
is in fact  periodic. Thus we can use Floquet theory to
conjugate $(\ref{al-ref3'})$ to a system with constant coefficients:

\begin{Lemma}\label{lemma-2}
There exist $A_2\in \mathrm{sl}(2,\mathbb{R})$, $F_2\in
C^\omega_{h_n}(\T^2,$ $\mathrm{sl}(2,\R))$ and $B_2\in
C^\omega_{h_n}(\T^2,\mathrm{PSL}(2,\R))$ with $\|B_2\|_{h_n}\leq
e^{q_{n+1}\varepsilon_n^{\frac{1}{4}}}$, $\deg B_2=0,$ such that
$B_2(\varphi)$ further conjugates the system (\ref{al-ref3}) to
\begin{eqnarray}\label{al-ref4}
\left\{ \begin{array}{l}\dot{\bx}=(A_2+F_2(\varphi))\bx \\
\dot{\varphi}=\omega=(1,\alpha)
\end{array} \right.
\end{eqnarray}
with estimates $|A_{2}|\leq e^{q_{n+1}\varepsilon_n^{\frac{1}{4}}}$, and $\|F_2\|_{ h_n}\leq
e^{-q_{n+1}\delta_n}$.
\end{Lemma}

\begin{pf}
Let $\varphi=(\varphi_1,\varphi_2)$ and $\phi=-p_n\varphi_1+q_n\varphi_2$, 
$\tau=-p_n+q_n\alpha$ and 
$$G(\phi)=\mathcal{T}_{\frac{q_{n+1}}{6}}F_1^{(re)}(\varphi)=
\sum_{ |l|< \frac{q_{n+1}}{6q_n}}F_1^{(re)} (-lp_n,lq_n)
e^{ 2\pi i l\phi }.$$
Consider the ODE 
\begin{equation}\label{period}\frac{d\bx}{dt}=G(\tau t)\bx,
\end{equation}which is a $\frac{1}{|\tau|}$-periodic linear equation.
Denote by $\Phi(t)$  the basic matrix solution of \eqref{period} with
$\Phi(0)=\id$. Thus by Floquet theory, \eqref{period} can be conjugated to $$\frac{d\bx}{dt}=A_2\bx$$ where $\Phi(\frac{2}{|\tau|})=e^{\frac{2}{|\tau|}A_2}$ by the conjugation  $\widetilde{B}(t)=e^{A_2t}\Phi(t)^{-1}$. 
Define 
$$B_2(\varphi_1,\varphi_2):=\widetilde{B}(-p_n\varphi_1+q_n\varphi_2/\tau),$$ then  $B_2(\varphi)$ conjugates \eqref{al-ref3'}
to 
\begin{eqnarray*}
\left\{ \begin{array}{l}\dot{x}=A_2x \\
\dot{\varphi}=\omega.
\end{array} \right.
\end{eqnarray*}
By the definition of rotation number and $(\ref{rot1})$, it follows that $$ \rho_f(\omega,\mathcal{T}_{\frac{q_{n+1}}{6}} F_1^{(re)}(\varphi))\leq \|\mathcal{R}_{\frac{q_{n+1}}{6}} F_1^{(re)}(\varphi)  \|_{C^0} \leq 2\epsilon_n e^{-\frac{q_{n+1}h_*}{6(1+\alpha)}} \ll \frac{1}{2q_{n+1}}.$$

Note that the constant matric $A_2$, the logarithm of  the matrix $\Phi(\frac{2}{|\tau|})$,  is not unique, which reflects the degree of the conjugacy $B_2$. The imaginary parts of eigenvalues of different $A_2$ (the rotation numbers) differ from each other  by  $k|\tau|$. We choose $A_2$ such that
\begin{equation}\label{rots1} \rho_f(\omega,A_2)= \rho_f(\omega,\mathcal{T}_{\frac{q_{n+1}}{6}} F_1^{(re)}(\varphi)) \ll \frac{1}{q_{n+1}}.\end{equation} 
At the same time, by Gronwall inequality, 
\begin{equation*} \|\Phi(t)\|_{C^0}\leq e^{t\|G\|_{C^0}}\leq e^{t\|\mathcal{T}_{\frac{q_{n+1}}{6}}F_1^{(re)}(\varphi)\|_{\frac{h_*}{1+\alpha}}}.\end{equation*}
Therefore, the constant matrix $A_2$ we choose satisfy 
$$| \mathfrak{spec}(A_2)| \leq \frac{|\tau|}{2}\max\{\|\mathcal{T}_{\frac{q_{n+1}}{6}}F_1^{(re)}(\varphi)\|_{\frac{h_*}{1+\alpha}},1\},$$
where $ \mathfrak{spec}(A_2)$ is the spectral radius of $A_2$.
Then the estimates on $B_2$ and $F_2$ follows from   Lemma $5.3$ of \cite{HoY} (see also Lemma 7.1 of \cite{HoY}) directly.
Note the equality \eqref{rots1}  implies the degree of $B_2(\varphi)$ is zero. \end{pf}

\begin{Remark} \label{reason}
Unfortunately Lemma \ref{lemma-2} can't be applied to
discrete cocycles. That's the main reason why   we need  the local embedding
theorem in \cite{YZ}.
\end{Remark}

Note that  $\deg B_2=0$ by Lemma \ref{lemma-2},    and so 
$\rho_f(\omega,A_2+F_2(\varphi) )=0$ by \eqref{rot-conj} and \eqref{rot1}.  Also since  we have assumed that $(\alpha,A)$ is subcritical, which implies that 
  $(\omega, A_2+F_2(\varphi))$
 is not uniformly hyperbolic. As a consequence,  by a change of variable with constant $\mathrm{SO}(2,\R)$ coefficients, we may assume that
$ A_2=d \cL$ with  $|d|= |A_2|\leq
 e^{q_{n+1}\varepsilon_n^{\frac{1}{4}}}$, one may consult \cite{YZ} (using Lemma 5.1) for similar proof. Finally,  $W_n^{'}(\varphi)=  B_1(\varphi)B_2(\varphi)H$
 will meet our needs where $H$ is defined as $\left(
\begin{array}{ccc}
\tilde \lambda & 0\\
0 &  \tilde{\lambda}^{-1}
 \end{array}\right) $ with  $ \tilde \lambda=e^{q_{n+1}\varepsilon_n^{\frac{1}{4}}}$. The proof of Proposition \ref{prop-hy} is thus finished.\qed\\

\section{Quantitative Aubry Duality}\label{qad}

In this section, we develop a  \textit{Quantitative version of Aubry duality}, which    is one of the key ingredients  in our proof. 
For simplicity,
in the rest of the paper, we assume $c^{'}$ to be
numerical constants.

%
%

\begin{Proposition}\label{prop-case1}
Let $\alpha\in\R\backslash \Q$,   $\lambda<1$,
$\cl \in \Z$, $E\in
\Sigma_{\lambda,\alpha}$, $0<2\pi h<-\ln \lambda.$
Then  \begin{eqnarray}\label{almost}
(-1)^\cl
B(x+\alpha)^{-1}A^{(\lambda,E)}(x)B(x)=
\id+F(x)
\end{eqnarray}
has  no solution $B(x)\in
C_{h}^\omega(\T,$ $\mathrm{PSL}(2,\R))$
with estimates $\|B\|_h\leq \varepsilon^{-1},$ $\|F\|_h\leq
\varepsilon^C$  if $\varepsilon$ is sufficiently small and
$C> c^{'} (\frac{\ln (4\lambda^{-1}+3)}{h})^2$.
\end{Proposition}

\begin{Remark}
We want to emphasize that Proposition \ref{prop-case1}  works for any $\alpha\in\R\backslash \Q$. The result is new even for  Diophantine frequencies  since it allows to get better control of the reducible parabolic cocycles, which will lead to
exponential lower bounds on the decay of the spectral
gaps \cite{LYZZ}. \end{Remark}

\begin{pf}

We will argue by contradiction. Let
\begin{eqnarray}\label{bf}B(x)= \left(
\begin{array}{ccc}
z_{11}(x) & z_{21}(x)\\
z_{12}(x) & z_{22}(x)
 \end{array}\right),\qquad F(x)=
\left(
\begin{array}{ccc}
\beta_{1}(x) & \beta_{2}(x)\\
\beta_{3}(x)& \beta_{4}(x)
 \end{array}\right). \end{eqnarray}
To fix ideas, we consider the case $(-1)^\cl=1$, the other case being
analogous.
By $(\ref{almost})$, we have 
 \begin{eqnarray}
\label{a-1} &&z_{11}(x)=z_{12}(x+\alpha)+z_{12}(x+\alpha)\beta_{1}(x)+z_{22}(x+\alpha)\beta_{3}(x),\\
\label{a-2}
&&z_{21}(x)=z_{12}(x+\alpha)\beta_{2}(x)+z_{22}(x+\alpha)\beta_{4}(x)+z_{22}(x+\alpha).
 \end{eqnarray}

Since the perturbation $F(x)$ may not be zero, the Fourier coefficients
$(\widehat{z}_{11}(k))$, $(\widehat{z}_{21}(k))\in l^2$ are not
necessarily to be the solutions of the almost Mathieu operator
\begin{equation}\label{mathieu-1}
\widehat{u}(k+1)+\widehat{u}(k-1) +2 \lambda^{-1} \cos 2\pi(
k\alpha)\widehat{u}(k)= E \lambda^{-1}\widehat{u}(k).
\end{equation}
However, $(\widehat{z}_{11}(k)) $ and $(\widehat{z}_{21}(k))$ are two approximate solutions of $(\ref{mathieu-1})$ if $F(x)$ is sufficiently small.

In fact, by $(\ref{almost})$, $z_{11}(x)$ satisfies the
following equation:
\begin{eqnarray}\label{block-red} (E-2\lambda
\cos2\pi (x))z_{11}(x)-z_{11}(x-\alpha)-z_{11}(x+\alpha)=f(x),\end{eqnarray}
where 
\begin{eqnarray} \label{pertur-f} f(x)=
z_{11}(x+\alpha)\beta_{1}(x)+
z_{21}(x+\alpha)\beta_{3}(x) 
-
z_{12}(x)\beta_{1}(x-\alpha)-z_{22}(x)\beta_{3}(x-\alpha),
\end{eqnarray}
with estimate $\|f\|_{h}\leq \|B\|_{h} \|F\|_{h}\leq
\varepsilon^{C-1}$.

Setting
$\A(x)=A^{(1/\lambda, E/\lambda)}(x)$, going to the Fourier coefficients, from $(\ref{block-red}),$ we have
\begin{eqnarray*}
(E -2\cos 2\pi (k\alpha))\widehat{z}_{11}(k) - \lambda ( \widehat{z}_{11}(k+1)+ \widehat{z}_{11}(k-1) )=
\widehat{f}(k).
\end{eqnarray*}
That is to say, $(\widehat{z}_{11}(k))$ is a solution of \begin{eqnarray*}\left(\begin{array}{c}
\widehat{z}_{11}(k+1) \\
\widehat{z}_{11}(k)
\end{array}\right)&= \A(k \alpha)  \left(\begin{array}{c}
\widehat{z}_{11}(k) \\
\widehat{z}_{11}(k-1)
\end{array}\right)-\left(\begin{array}{c}
\lambda^{-1}
\widehat{f}(k) \\
0
\end{array}\right).
\end{eqnarray*}
Intuitively, one sees that $(\widehat{z}_{11}(k))$ is an
approximating solution of
\begin{eqnarray}\label{mathieu}\left(\begin{array}{c}
\widehat{u}_{1}(k+1) \\
\widehat{u}_{1}(k)
\end{array}\right)&=&
 \A(k \alpha)  \left(\begin{array}{c}
\widehat{u}_{1}(k) \\
\widehat{u}_{1}(k-1)
\end{array}\right)
\end{eqnarray} if  $\lambda^{-1} \widehat{f}(k)$ is very small.

Similarly, by $(\ref{almost})$, $(\widehat{z}_{21}(k))$ satisfies
the following equation:
\begin{eqnarray*}
(E -2\cos 2\pi (k\alpha))\widehat{z}_{21}(k) - \lambda ( \widehat{z}_{21}(k+1)+ \widehat{z}_{21}(k-1) )=
\widehat{g}(k).
\end{eqnarray*}
where  $\widehat{g}(k)$'s  are  Fourier coefficients of 
\begin{eqnarray} \label{pertur-g} g(x)=
z_{11}(x+\alpha)\beta_{2}(x)+
z_{21}(x+\alpha)\beta_{4}(x) 
-
z_{12}(x)\beta_{2}(x-\alpha)-z_{22}(x)\beta_{4}(x-\alpha).
\end{eqnarray}
It is easy to see that  $\|g\|_{h}\leq \|B\|_{h}\|F\|_{h}\leq
\varepsilon^{C-1}$.\\

The proof of Proposition \ref{prop-case1} will be decomposed into the following lemmas. First, we prove: 

\begin{Lemma}\label{z1-estimate}
Let $z_{11}(x)$ be the function defined in $(\ref{bf})$ satisfying \eqref{almost}. Then the following estimates hold:
\begin{enumerate}
\item let $\|z_{11}(x)\|_{L^2}= (\int |z_{11}(x)|^2 dx)^{\frac{1}{2}}$, then 
 \begin{equation} \label{z1-estimate-1}\|z_{11}\|_{L^2} \geq \frac{1}{4}\|B\|_{C^0}^{-1} \geq  \frac{1}{4} \varepsilon,
\end{equation}
\item  there exists $s\in [0, \frac{3 \ln \varepsilon^{-1}}{2\pi h}]$, such that
\begin{equation}\label{z1-estimate-2}|\widehat{z}_{11}(s)|> \varepsilon^{2}.\end{equation}
\end{enumerate}
\end{Lemma}

\begin{pf}
 Write $$u=\left(\begin{array}{c} z_{11}(x) \\
z_{12}(x)
\end{array}\right), \qquad v=\left(\begin{array}{c} z_{21}(x) \\
z_{22}(x)
\end{array}\right).$$ Then
$\|u\|_{L^2}\|v\|_{L^2}>1$ since $\det B(x)=1.$ It follows  that
$$\|u\|_{L^2}> \frac{1}{\|v\|_{L^2}}>\frac{1}{\|B\|_{C^0}}.$$
 By $(\ref{a-1})$, we have
$$z_{12}(x)=z_{11}(x-\alpha)-z_{12}(x)\beta_{1}(x-\alpha)-z_{22}(x)\beta_{3}(x-\alpha).$$
Therefore,
\begin{equation*}2\|z_{11}\|_{L^2}+2\|B\|_{C^0}\|F\|_{C^0}>\frac{1}{\|B\|_{C^0}}.\end{equation*}
This estimate implies $(\ref{z1-estimate-1})$ since $\|B\|_{C^0}\|F\|_{C^0}$ is small enough. The first
statement is now proved.

The second statement follows directly from the first one. Since
$|\widehat{z}_{11}(k)|\leq  \varepsilon^{-1}e^{- 2\pi |k|h}$,  then
combining with $(\ref{z1-estimate-1})$, we have for $K= \frac{3\ln
\varepsilon^{-1}}{ 2\pi h}$,
 $$
\sum_{|k|\leq  K}|\widehat{z}_{11}(k)|^2 \geq \|z_{11}\|_{L^2}^2
-\sum_{|k|> K}|\widehat{z}_{11}(k)|^2 > \frac{1}{20}\varepsilon^{2}.$$ Let $ |\widehat{z}_{11}(s)|= \max_{|k|\leq K}
|\widehat{z}_{11}(k)|$. It follows that $|\widehat{z}_{11}(s)|>
\varepsilon^{2}$. Notice that $z_{11}(x)$ is real analytic, which
implies $\overline{\widehat{z}}_{11}(s)=\widehat{z}_{11}(-s)$,  then one can always select $s\in [0,K] $ such that (\ref{z1-estimate-2}) holds. We thus finish the whole proof.\end{pf}

In the following, $s\geq 0$ is chosen so that (\ref{z1-estimate-2}) is satisfied. Denote $$\widetilde{D}_s:=\det \left(
\begin{array}{ccc}
 \widehat{z}_{11}(s) &  \widehat{z}_{21}(s)\\
\widehat{z}_{11}(s+1) &  \widehat{z}_{21}(s+1)\end{array} \right).$$
 With the
help of Lemma \ref{z1-estimate}, we have:

\begin{Lemma}\label{det-estimate}
  $\widetilde{D}_s \ge
\varepsilon^{\kappa}$ with $\kappa= \frac{ 2\pi h}{\ln(4\lambda^{-1}+3)+2\pi h}C$.
\end{Lemma}

\begin{pf} We prove this lemma by contradiction. If $\widetilde{D}_s <
\varepsilon^{\kappa},$  then by the fact that 
\begin{eqnarray}
\label{ortho} \left(
\begin{array}{ccc}  \widehat{z}_{21}(s) \\ \widehat{z}_{21}(s+1) \end{array}
\right)
= \left(
\begin{array}{ccc} C_s \widehat{z}_{11}(s) \\ C_s  \widehat{z}_{11}(s+1) \end{array}
\right)
+
\frac{\widetilde{D}_s}{\sqrt{|\widehat{z}_{11}(s)|^2+ |\widehat{z}_{11}(s+1)|^2}} \left(
\begin{array}{ccc} 
-\overline{\widehat{z}_{11}(s+1)} \\ \overline{\widehat{z}_{11}(s)} \end{array}
\right)
\end{eqnarray}
one sees that
\begin{eqnarray}\label{initial}
\left|\left(
\begin{array}{ccc}
 \widehat{z}_{21}(s) -C_s\widehat{z}_{11}(s)\\
\widehat{z}_{21}(s+1) -C_s\widehat{z}_{11}(s+1)\end{array}
\right)\right| \leq \widetilde{D}_s< \varepsilon^{\kappa},
\end{eqnarray}
which means that the orthogonal projection of $(\widehat{z}_{21}(s),
\widehat{z}_{21}(s+1))^{T}$ to the vector $(\widehat{z}_{11}(s),
\widehat{z}_{11}(s+1))^{T}$ is small. By $(\ref{ortho})$, we have
$$C_s=\frac{ \widehat{z}_{21}(s)+D_s \overline{\widehat{z}_{11}(s+1)} /
\sqrt{|\widehat{z}_{11}(s)|^2+ |\widehat{z}_{11}(s+1)|^2} }{
\widehat{z}_{11}(s)}.$$  Then by Lemma \ref{z1-estimate}(2), we obtain
$|C_s|<\varepsilon^{-3}.$

Denote $$\widetilde{p}_k = \left(\begin{array}{c}
\lambda^{-1}\widehat{g}(k)-\lambda^{-1} C_s
\widehat{f}(k) \\
0
\end{array}\right),\qquad \widetilde{y}_k=\left(\begin{array}{c}
 \widehat{z}_{21}(k+1) -C_s \widehat{z}_{11}(k+1)\\
\widehat{z}_{21}(k) -C_s \widehat{z}_{11}(k)
\end{array}\right),$$
where $\widehat{f}(k),\widehat{g}(k)$ are Fourier coefficients of
$f(x),g(x)$
defined in $(\ref{pertur-f}),$ $(\ref{pertur-g})$ respectively. Since both $(
\widehat{z}_{11}(k))$ and $( \widehat{z}_{21}(k))$ are  approximate
solutions of  the almost Mathieu operator $(\ref{mathieu})$, then we
have the following relationship
\begin{equation}\label{iter-1}
\widetilde{y}_k=\A(k\alpha)
\widetilde{y}_{k-1}+\widetilde{p}_{k}.
\end{equation}  It follows that
\begin{equation*}
\widetilde{y}_k= \A^{k-s}((s+1)\alpha) \widetilde{y}_s+
\sum_{j=s+1}^{k}\A^{k-j}((j+1)\alpha)\widetilde{p}_j,
\end{equation*}
where  $\A^k(x)= \A(x+(k-1)\alpha)\cdots \A(x).$ By $(\ref{initial})$, we have $|
\widetilde{y}_s|\leq \varepsilon^{\kappa}.$  Therefore for any $k$, we have
\begin{eqnarray}
\label{diffence}|\widetilde{y}_k|&\leq&  | \A^{k-s}(\alpha) \widetilde{y}_s|+  \sum_{j=s+1}^{k}\|\A^{k-j}((j+1)\alpha)\||\widetilde p_j|     \leq  \varepsilon^{\kappa}(4\lambda^{-1}+3)^{|k|+s},
\end{eqnarray}
here we use the fact that if $E\in \Sigma_{\lambda,\alpha}$, then $\lambda^{-1} E\in  \Sigma_{\lambda^{-1},\alpha}$ by Aubry duality \cite{Au}, which implies that  \footnote{One can use Theorem \ref{bj-formula} to optimize the lower bound of the constant $C$ appeared in Proposition \ref{prop-case1}.}
$\|\A(x)\| \leq 4\lambda^{-1}+3$.

Since $\det B(x)=1$, we have
$$
\det\left(
\begin{array}{ccc}
z_{11}(x) & z_{21}(x)-C_sz_{11}(x)   \\
z_{12}(x) & z_{22}(x)-C_sz_{12}(x)
 \end{array}\right)=1.
$$
Let $L= \frac{6 \ln \varepsilon^{-1}}{ \pi h}$, it is easy to see
 $$\| \mathcal {R}_L(z_{21}(x)-C_sz_{11}(x)
)\|_{\frac{h}{2}} \leq \varepsilon^{-4}e^{- \pi  Lh}= \varepsilon^2, $$
then by $(\ref{a-1})$,$(\ref{a-2})$, we get
\begin{eqnarray}
\label{det}&&\det\left(
\begin{array}{ccc}
z_{11}(x) & \mathcal {T}_L(z_{21}(x)-C_s z_{11}(x))   \\
z_{11}(x-\alpha) & \mathcal {T}_L(z_{21}(x-\alpha)-C_s
z_{11}(x-\alpha))
 \end{array}\right)\\
 \nonumber &\geq& \det B(x)- 2 \|B\|_{h}\| \mathcal
{R}_L(z_{21}(x)-C_s z_{11}(x)
)\|_{\frac{h}{2}}-2|C_s|\|B\|_{h}^2\|F\|_{h} \\
\nonumber&\geq& 1- 2\varepsilon-2\varepsilon^{C-5}> \frac{1}{2}.
\end{eqnarray}

 On the other
hand, by the selection of $s$ and $(\ref{diffence})$,  we have
\begin{eqnarray*} \| \mathcal {T}_L(z_{21}(x)-C_s z_{11}(x)
)\|_{\frac{h}{2}}  <  \sum_{|k|\leq L} \varepsilon^{\kappa}
       (4\lambda^{-1}+3)^{|k|+s} e^{\pi  |k| h}
<   \varepsilon^{\kappa-6- \frac{15\ln (4\lambda^{-1}+3)}{ 2\pi h}} <
\varepsilon^2,
       \end{eqnarray*}
the last inequality holds since we select $C> c^{'} (\frac{\ln
(4\lambda^{-1}+3)}{h})^2$. Together with the fact that the norms of
$z_{11}(x), z_{12}(x)$ are bounded by $\|B\|_{h}$, then
it contradicts with $(\ref{det})$.
\end{pf}

Let $(\widehat{u}_1(k))$ be the solution of the almost Mathieu operator
 $(\ref{mathieu})$ with
 initial datum
$\widehat{u}_1(s)=\widehat{z}_{11}(s)$,
$\widehat{u}_1(s+1)=\widehat{z}_{11}(s+1)$,  and let
$(\widehat{u}_2(k))$ be the solution of  $(\ref{mathieu})$ with
$\widehat{u}_2(s)=\widehat{z}_{21}(s)$,
$\widehat{u}_2(s+1)=\widehat{z}_{21}(s+1)$. Define  the Wronskian determinant
$$\widetilde{D}_k':=\det \left(
\begin{array}{ccc}
 \widehat{u}_{1}(k) &  \widehat{u}_{2}(k)\\
\widehat{u}_{1}(k+1) &  \widehat{u}_{2}(k+1)\end{array} \right),$$
 which is in fact a constant, i.e. $\widetilde{D}_k'= \widetilde{D}'$ for all $k$.
Note  that $(\widehat{u}_{1}(k))$ and
$(\widehat{u}_{2}(k))$ are linearly independent since otherwise $\widetilde{D}_s=\widetilde{D}_{s}^{'}=0$, which contradicts to Lemma \ref{det-estimate}. The following lemma measures quantitatively the linear independence of the sequences  $(\widehat{u}_{1}(k))$ and
$(\widehat{u}_{2}(k))$:

\begin{Lemma}\label{det-estimate-2}
For $ k\in K_n =[\frac{2\ln \varepsilon^{-1}}{\pi h}, \frac{C\ln
\varepsilon^{-1}}{\ln (4\lambda^{-1}+3)+ 2\pi h}],$ we have
 $$|\widetilde{D}_k^{'}|\leq
8 \varepsilon^{-2}e^{-4\pi |k|h}.$$
\end{Lemma}

\begin{pf}
 First we remark that,  from our selection way of $C$,  it is  obvious that  $K_n\neq
\emptyset$. By symmetry,  it suffices  for us to prove that  $$|\widehat{u}_{1}(k)|\leq
2 \varepsilon^{-1}e^{-2\pi  |k|h}.$$ Denote $$p_k =
\left(\begin{array}{c} -\lambda^{-1}
\widehat{f}(k) \\
0
\end{array}\right), \qquad y_k=\left(\begin{array}{c}
\widehat{u}_{1}(k+1) - \widehat{z}_{11}(k+1)\\
\widehat{u}_{1}(k)-\widehat{z}_{11}(k)
\end{array}\right),$$ then by the similar argument as in Lemma
\ref{det-estimate}, we know
\begin{equation*}
y_k= \A^{k-s}((s+1)\alpha) y_s+
\sum_{j=s+1}^{k}\A^{k-j}((j+1)\alpha)p_j.
\end{equation*} Since the initial datum of
$\widehat{z}_{11}(k)$ and
 $\widehat{u}_{1}(k)$ are equal, i.e.
$y_s=0$, then for any $k>s$,
\begin{eqnarray*}
|y_k|&\leq&\sum_{j=s+1}^{k}\|\A^{k-j}((j+1)\alpha)\||p_j|\\
       &\leq& \varepsilon^{C-1}\sum_{j=s+1}^{k}(4\lambda^{-1}+3)^{k-j}\lambda^{-1} e^{- 2\pi  |j|
       h}\leq  \varepsilon^{C-1} e^{k\ln(4\lambda^{-1}+3)}.
\end{eqnarray*}

On the other hand,  $\widehat{z}_{11}(k)$ decays exponentially  when $k$ is large enough since $\widehat{z}_{11}(k)$'s are  the Fourier
coefficients of the analytic function $z_{11}(x)$. If
we take $k\geq \frac{2\ln \varepsilon^{-1}}{\pi h}$, then
$$
|\widehat{z}_{11}(k)| \leq \varepsilon^{-1} e^{-2\pi  |k|h}<
\varepsilon^3,
$$
thus when $k\in K_n,$  $\widehat{u}_{1}(k)$ also decays
exponentially with estimate
\begin{eqnarray*}
|\widehat{u}_{1}(k)|\leq |y_k|+|\widehat{z}_{11}(k)|
\leq \varepsilon^{C-1} e^{k\ln(4\lambda+3)}+\varepsilon^{-1}
e^{- 2\pi  |k|h}\leq 2 \varepsilon^{-1} e^{-2\pi  |k|h},
\end{eqnarray*}
we therefore finish the proof.
\end{pf}

We can now finish the  proof  of Proposition \ref{prop-case1}. 
Since $(\widehat{u}_{1}(k))$,
$(\widehat{u}_{2}(k))$  are  two linearly independent  solutions of
the almost Mathieu operator $(\ref{mathieu})$, then by
Liouville's theorem, we have $\widetilde{D}_{k}^{'}=\widetilde{D}_{s}^{'}=\widetilde{D}_s.$ Take
$k=\frac{C\ln \varepsilon^{-1}}{\ln (4\lambda^{-1}+3)+2\pi  h}.$  By Lemma
\ref{det-estimate-2} and our selection of $\kappa$ and $C$, we have
$$\widetilde{D}_s=\widetilde {D}_{k}^{'}<8
\varepsilon^{\frac{4\pi  Ch}{\ln (4\lambda^{-1}+3)+2\pi  h}-2}< \frac{1}{2}\varepsilon^{\kappa},$$
which contradicts with Lemma \ref{det-estimate}.
\end{pf}

\section{Generalized Moser-P\"oschel Argument}\label{section5}

%

We  briefly outline the ideas of the proof.  Assume that
$0<\lambda<1$, $E$ is in the spectrum  and 
$\cl=2\rho_f(\alpha,A^{(\lambda,E)})-k_0\alpha \in \Z.$ Thanks to
Theorem \ref{prop-main} (resp. Theorem \ref {prop-main-dio})
and Proposition \ref{prop-case1},
$(\alpha, A^{(\lambda,E)} )$ can be conjugated  to a parabolic cocycle with
an arbitrarily small perturbation, and moreover the off-diagonal
element of the parabolic matrix has good
quantitative estimates. If $(\alpha, (-1)^\cl A^{(\lambda,E)} )$ is
reducible to a parabolic cocycle $\id+d \cL$ with $d \neq 0$,
by the classical Moser-P\"oschel argument \cite{MP84}, one can prove
that  $(\alpha, A^{(\lambda,E+\tau)} )$ is uniformly hyperbolic
for sufficient small $\tau$ (the sign of 
$\tau$ depends on the sign of the off-diagonal
element of the parabolic matrix $D$).
This implies the openness of the gap.  This is the (previously known)
argument for the Diophantine case, when we can apply
Theorem \ref {prop-main-dio}.
As mentioned in Section \ref {1.2}, here we develop
a  \textit{generalized Moser-P\"oschel argument} that can be applied for
non-Diophantine $\alpha$ as well.  It shows that
$(\alpha, A^{(\lambda,E+\tau)} )$ is uniformly hyperbolic with
$2 \rho_f(\alpha,A^{(\lambda,E+\tau)})=k_0\alpha \mod  \Z$ for some
positive or negative  $\tau$.  Openness of the corresponding
gap follows by the trict
monotonicity of the integrated density
of states restricted to the spectrum.
We stress that the proof of uniformly hyperbolicity of
$(\alpha, A^{(\lambda,E+\tau)} )$ makes use of
the  delicate estimates on the conjugacies  and the errors  of the perturbation as showing in Theorem  \ref{prop-main}.  \\

\noindent
\textbf{Proof of Theorem \ref{dry}:}  By  Aubry duality \cite{GJLS},   it suffices to consider
$H_{\lambda,\alpha,x}$ for $0<\lambda<1$. From now on, we
fix $k_0\in\Z$ and assume that $E_{k_0}$ is in the spectrum  with
$\cl=2\rho_f(\alpha,A^{(\lambda,E_{k_0})})-k_0\alpha \in \Z.$
By Theorem \ref{bj-formula}, the cocycle  $(\alpha,
A^{(\lambda,E_{k_0})})$ is subcritical in the
region $|\mathfrak{I}x|< \frac{-\ln\lambda}{2\pi }$.

In the case $\beta(\alpha)=0$, we apply Theorem \ref {prop-main-dio} and
Proposition \ref {prop-case1} to concluded that
$(\alpha,(-1)^\cl A^{(\lambda,E_{k_0})})$ is conjugated by some analytic
$B$ to $\id+d \cL$ with
$d \neq 0$.  For definiteness, assume that $d>0$, the other case being
analogous.  For $0 \leq a<b <\infty$, let
$\cC_{a,b} \in \P\R^2$ be the cone of directions of vectors
$(x,y) \in \R^2$ with $a x \leq y \leq b x$.  Then
$\cC_{0,1}$ is taken to $\cC_{0,(1+d)^{-1}} \subset \cC_{0,1}$.
Then for some $C>0$ (depending most crucially on the size of $B$) and for
$-\frac {C^{-1} d} {1+d}<\tau<0$, $$\tilde A_\tau(x)= (-1)^\cl B(x+\alpha)^{-1}  
A^{(\lambda,E_{k_0}+\tau)}
B(x)$$ sends $\cC_{0,1}$ inside $\cC_{0,(1+d)^{-1}+C \tau} \subset
\cC_{0,1}$ for some $C>0$ (depending most crucially on the size of $B$). 
This shows that
$\rho_f(\alpha,\tilde A_\tau)=0$ and hence
$\rho_f(\alpha,A^{(\lambda,E_{k_0}+\tau)})=
\rho_f(\alpha,A^{(\lambda,E_{k_0})})$.  In particular the size of the
corresponding gap is at least $\frac {C^{-1} d} {1+d}$.  This argument is
further developed in \cite {LYZZ} to yield explicit bounds.

We now assume $\beta(\alpha)>0$ and apply Theorem \ref{prop-main}, to
conclude that 
there exist a subsequence of $q_n$, $B_n(x)\in C_{h_n}^\omega(\T,\mathrm{PSL}(2,\R))$ with
$\|B_n\|_{h_n}\leq e^{2q_{n+1}\varepsilon_n^{\frac{1}{4}}}$, such
that
\begin{eqnarray}\label{almost-1}
(-1)^\cl B_n(x+\alpha)^{-1}A^{(\lambda,E_{k_0})}(x)B_n(x)=
\id+d_n \cL+F_n(x)
\end{eqnarray}
with estimate $\|F_n\|_{h_n}\leq e^{-q_{n+1}\delta_n}$.  And one can choose $n=n_{k_0}(\lambda)$ large enough such that
$$c\delta_n > c^{'}\varepsilon_n^{\frac{1}{4}}(\frac{\ln (4\lambda^{-1}+3)}{h_n})^2.$$
We can now
apply Proposition \ref{prop-case1} to bound $d_n$ away from zero,
specifically $ |d_n|>  e^{-q_{n+1}\varepsilon_n^{\frac{1}{8}}}.$

In the following, we generalize the Moser-P\"oschel argument to
prove the openness of the gap. We first prove the following:
\begin{Proposition}\label{mp-argument}
Suppose that $\alpha\in \R\backslash
\Q$, and the cocycle $(\alpha,A^{(\lambda,E_{k_0})})$ satisfies (\ref
{almost-1}) with estimates given in Theorem \ref{prop-main} and $$ 
 e^{-q_{n+1}\varepsilon_n^{\frac{1}{8}}}
\leq |d_n|<e^{-q_{n+1}\varepsilon_n^{\frac{1}{4}}}.$$ Then the
cocycle $(\alpha,A^{(\lambda,E_{k_0}+\tau)})$ is uniformly
hyperbolic with
$2\rho_f(\alpha,A^{(\lambda,E_{k_0}+\tau)}(x))=k_0\alpha \mod  \Z$ for all $\tau$ satisfying $d_n\tau<0$ and
$$|\tau|\in I_n =( e^{-\frac{q_{n+1}
\delta_n}{2}}, e^{-4q_{n+1}\varepsilon_n^{\frac{1}{8}}}).$$  
\end{Proposition}

\begin{pf}
For definiteness we will consider the case $(-1)^\cl=1$, the other case
being analogous.
If we write $B_n(x)$ and $F_n(x)$  as in 
$(\ref{bf})$, then one sees that $(\ref{a-1})$ still holds, while
 $(\ref{a-2})$ is replaced by
  \begin{equation}\label{a-2'}
z_{21}(x)=d_n
z_{12}(x+\alpha)+z_{12}(x+\alpha)\beta_{2}(x)+z_{22}(x+\alpha)\beta_{4}(x)+z_{22}(x+\alpha).
\end{equation}
Straightforward  computation shows that, the transformation
given by $(0,B_n(\cdot))$ conjugates the cocycle
$(\alpha,A^{(\lambda,E_{k_0}+\tau)}(x))$   to $(\alpha,
A_1(\cdot))$ with
\begin{eqnarray*}
A_1 (x)  =\left(
\begin{array}{ccc}
1+ \beta_{1}(x) + \tau (z_{11}(x) z_{22}(x+\alpha)) &
d_n+ \beta_{2}(x) + \tau ( z_{21}(x) z_{22}(x+\alpha)) \\
\beta_{3}(x) -\tau( z_{11}(x) z_{12}(x+\alpha)) &1+
\beta_{4}(x) - \tau( z_{21}(x) z_{12}(x+\alpha))
\end{array}\right).
\end{eqnarray*}
By $(\ref{a-1}),(\ref{a-2'})$,  further
computation leads to
\begin{eqnarray*}
z_{11}(x) z_{22}(x+\alpha)&=& z_{11}(x)
z_{21}(x) - d_n z_{11}^2(x)- d_nz_{11}(x)(
z_{12}(x+\alpha)\beta_{1}(x)+ \\ && z_{22}(x+\alpha)\beta_{3}(x))
 - z_{11}(x)(
z_{12}(x+\alpha)\beta_{2}(x)+z_{22}(x+\alpha)\beta_{4}(x)
),
\end{eqnarray*}
 and the other elements in the matrix $A_1(x)$ can be computed  similarly.

Write $A_1(x) =\id+ d_n \cL+M_n(x)+P_n(x)$, 
where $$ M_n(x) = \left(
\begin{array}{ccc}
 \tau\left( z_{11}(x) z_{21}(x)- d_n z_{11}^2(x) \right)
    &  \;\;\;  \tau\left(-d_n z_{11} (x)z_{21}(x) + z_{21}^2(x)\right) \\
-\tau z_{11}^2(x)  & \;\;\;  - \tau z_{11}(x) z_{21}(x)
\end{array}
\right).$$
Then  $P_n(x)$ is of the size     $$P_n\approx O( |\tau| \|B\|_{h_n}^2 \|F_n\|_{h_n}+
\|F_n\|_{h_n}).$$ 

To prove  $(\alpha, A_1)$ is actually uniformly hyperbolic for
$|\tau|\in I_n$ with  $d_n\tau<0$,  we will prove that $(\alpha,\id+d_n \cL+M_n(\cdot))$ is uniformly hyperbolic and the uniform hyperbolicity will not be destroyed by $P_n(\cdot)$. 
To investigate the hyperbolicity of $(\alpha,\id+d_n \cL+M_n(\cdot))$, we need  do one more  step of Newtonian iteration to eliminate the non-resonant terms of $M_n(\cdot)$. 

Let $$\Lambda=\{k\in
\mathbb{Z}:
 \|k\alpha\|_{\R/\Z}\geq \frac{1}{7q_{n+1}}\}, $$
and define
\begin{eqnarray*}
\widetilde{\mathfrak{B}}_{h_n}^{(nre)}=\left\{Y \in
C^\omega_{h_n}(\T,\mathrm{sl}(2,R))  \Big| Y=\sum_{k\in \Lambda }
\left(\begin{array}{ccc}
  \widehat{y}_{11}(k) &  \widehat{y}_{21}(k)\\
 \widehat{y}_{12}(k) &  -  \widehat{y}_{11}(k)
 \end{array}\right)e^{ 2\pi i\langle k,x \rangle}
\right\}.
\end{eqnarray*}
Denote by $Ad_{A}B$ the matrix $A^{-1} B A$ and  $P^{(nre)}$ the projection operator on $\widetilde{\mathfrak{B}}_{h_n}^{(nre)}$. If $M\in \widetilde{\mathfrak{B}}_{h_n}^{(nre)},$ 
then the homological equation
$$ Ad_{\id+d_n \cL}Y (x+\alpha)-Y(x) =M(x),$$
which is 
\begin{eqnarray*}
y_{12}(x+\alpha)-y_{12}(x)&=&M_{12}(x),\\
y_{11}(x+\alpha)-y_{11}(x)&=&M_{11}(x)+d_n y_{12}(x+\alpha) ,\\
y_{21}(x+\alpha)-y_{21}(x)&=&M_{21}(x)+d_n^2
y_{12}(x+\alpha)-2d_n y_{11}(x+\alpha),
\end{eqnarray*}
has a unique solution with estimate
$$\|Y\|_{h_n}\le cq_{n+1}^3\|M\|_{h_n}.$$

Notice that if $|\tau|<e^{-8q_{n+1} \varepsilon_n^{\frac{1}{4}}}$, then
$M_n(\cdot)$ is of the size $$\|M_n\|_{h_n}\leq c^{''}
|\tau|\|B_n\|_{h_n}^2\leq c^{''} e^{-4q_{n+1}
\varepsilon_n^{\frac{1}{4}}}\ll \frac{c}{q_{n+1}^6}.$$
Thus we can apply one step of Newtonian iteration, and get $Y(x)$  such that
$$e^{-Y(x+\alpha)}(\id+d_n \cL+ M_n(x))e^{Y(x)}= \id+d_n \cL+ {M_n}^{(re)}(x)+ R_n(x),$$
where ${M_n}^{(re)}= M_n- P^{(nre)} M_n$, $R_n(x)$ is of size 
 \begin{eqnarray*}
 R_n &\approx& O( \|Y\|_{h_n}^2 +  \|M_n\|_{h_n}^2+  \|Y\|_{h_n} \|M_n\|_{h_n} )\\ 
 &\approx& O( (q_{n+1}^3|\tau |
\|B_n\|_{h_n}^2)^{2} )  \approx O( |\tau|^{2} e^{8q_{n+1} \varepsilon_n^{\frac{1}{4}}}) 
\end{eqnarray*}
 Moreover, we have estimates\begin{eqnarray*}
\|Y\|_{h_n} &\leq& cq_{n+1}^3 \|M_n\|_{h_n} \leq e^{-q_{n+1}
\varepsilon_n^{\frac{1}{4}}/2},\\
\|{M_n}^{(re)}\|_{h_n} &\leq& 2 \|M_n\|_{h_n} \leq |\tau|
\|B\|_{h_n}^2
\end{eqnarray*}
For the details of the iteration step and the estimates,  the reader is refered to  Lemma 3.1 of \cite{HoY}.

The following  lemma on the small divisors is needed to prove that the lower order resonant terms in $M_n^{(re)}(\cdot)$ 
has some structure.

\begin{Lemma}\label{divisor}
$\|k\alpha\|_{\R/\Z}\geq \frac{1}{7q_n} $ holds for any $k\in \mathbb{Z}$ satisfing
$|k|\leq \frac{1}{6}q_{n+1}$
and  $k\neq l q_n.$

\end{Lemma}

\begin{pf}
See Lemma 4.1 \cite{HoY}.
\end{pf}

By Lemma \ref{divisor}, one knows that, the truncation of  ${M_n}^{(re)}$ up to the order  $\frac{q_{n+2}}{6}$ 
has the form
$$ \mathcal {T}_{\frac{q_{n+2}}{6}}{M_n}^{(re)}=\sum_{ k=l q_{n+1}}\widehat{{M_n}}(k)e^{ 2\pi  i<k,x>},$$
and then 
$$ \|{M_n}^{(re)}-[M_n]\|_{h_n-6\delta_n}\leq |\tau|
\|B_n\|_{h_n}^2e^{-q_{n+1}\delta_n}.$$  Thus, $(0, e^{Y(\cdot)})$ conjugates the cocycle
$(\alpha, A_1)$ to $(\alpha, A_2)$ with
$$A_2=
\left(
\begin{array}{ccc}
1 + \tau\left( [z_{11} z_{21}] - d_n [z_{11}^2] \right)
    &  \;\;\; d_n + \tau\left(-d_n [z_{11} z_{21}] + [z_{21}^2]\right) \\
-\tau [z_{11}^2]  & \;\;\; 1 - \tau [z_{11} z_{21}]
\end{array}
\right)+ \widetilde{M}
$$
where
$$ \widetilde{M}(x)={M_n}^{(re)}(x)-[M_n]+R_n(x)+ e^{-Y(x+\alpha)}P_n(x)e^{Y(x)}  $$
with  estimate 
\begin{eqnarray*}
\| \widetilde{M}\|_{h_n} &\leq& 
 |\tau| \|B_n\|_{h_n}^2 e^{-q_{n+1}\delta_n}+\| R_n\|_{h_n} +  
\|F_n\|_{h_n}\\&\leq& |\tau|^{2} e^{8q_{n+1} \varepsilon_n^{\frac{1}{4}}}+2 e^{-q_{n+1}\delta_n}
\leq 3 |\tau|^{2} e^{8q_{n+1} \varepsilon_n^{\frac{1}{4}}},
   \end{eqnarray*}
the inequality holds since by our assumption $|\tau|\in I_n$.

 The
$x$-independent part of the cocycle $(\alpha,A_2)$, i.e.,  $(\alpha, A_2-\widetilde M)$,  is hyperbolic if
$d_n\tau<0$. In fact, the  trace of $A_2-\widetilde M$ is $2 -d_n\tau [z_{11}^2] $ which is  greater than 2 since $z_{11}(x)$ is 
nonzero.  

To prove the hyperbolicity of $(\alpha, A_2)$, we have to prove that the hyperbolicity of $(\alpha, A_2-\widetilde M)$
will be preserved when the perturbation $\widetilde M$ is added.
We first check the hyperbolicity of the
constant cocycle  $(\alpha, A_2-\widetilde M)$.  Since  $z_{11}\in
C_h^\omega(\T,\R)$ is real analytic, we have\footnote{The proof of (\ref{111})  is exactly same as that of Lemma \ref{z1-estimate}, the only diffence is to replace (\ref{a-2}) by (\ref{a-2'}).}
\begin{equation}
\label{111} [z_{11}^2]= \sum |\widehat{z}_{11}(k)|^2=\|z_{11}\|_{L^2}^2 \geq c^{'} \|B_n\|_{C^0}^{-2}\geq 
c^{'} e^{-4q_{n+1} \varepsilon_n^{\frac{1}{4}}} .
\end{equation} It follows that
the hyperbolicity of the constant cocycle $(\alpha,A_2-\widetilde M)$ is bounded from below by
$$ |d_n\tau [z_{11}^2]| \geq |\tau| e^{-q_{n+1}\varepsilon_n^{\frac{1}{8}}} e^{-4q_{n+1}\varepsilon_n^{\frac{1}{4}}}. $$
Note that $\| \widetilde{M}\|_{h_n} \leq 3 |\tau|^{2} e^{8q_{n+1} \varepsilon_n^{\frac{1}{4}}}$ if $|\tau|\in I_n$, then 
it is easy to see that the cocycle  $(\alpha, A_2(x))$ is uniformly hyperbolic when
$d_n\tau<0$, one can consult  Proposition 18 of \cite{P2} for similar details. 
Furthermore, by the definition of the rotation number, it is easy to check that $\rho_f(\alpha,A_2(x))=0$. 

By the construction, one has 
$$e^{-Y(x+\alpha)}B_n(x+\alpha)^{-1}A^{(\lambda,E_{k_0}+\tau)}(x)B_n(x)e^{Y(x)}=A_2(x).$$ It follows that $(\alpha, A^{(\lambda,E_{k_0}+\tau)})$ is uniformly hperbolic when
$d_n\tau<0$ and $|\tau|\in I_n$.
Since the transformation $e^{Y(\cdot)}$ is close to constant, we have $\deg(e^{Y(\cdot)})=0$.  Also since $\deg B_n=k_0$, then by $(\ref{rot-conj})$, we have
$2\rho_f(\alpha,A^{(\lambda,E_{k_0}+\tau)}(x))=k_0\alpha \quad \text{mod}\  \Z$ for any $|\tau|\in I_n$ and $d_n\tau<0$.
The proof of Proposition \ref{mp-argument} is thus finished.\end{pf}

Now we are in the position to finish the proof of Theorem \ref{dry}.  Without loss of generality, we assume $d_{n}<0$.  By
 Proposition \ref{mp-argument}, we know that for $\tau\in I_n$,
  \[2\rho_f(\alpha,A^{(\lambda,E_{k_0}+\tau)}(x))=2\rho_f(\alpha,A^{(\lambda,E_{k_0})}(x))=k_0\alpha \quad \text{mod} \, \Z.\]
The rotation number is  strictly
monotonic restricted to the spectrum, then it follows that $2\rho_f(\alpha,A^{(\lambda,E')}(x))=k_0\alpha \quad \text{mod} \, \Z$ for
any $E'\in [E_{k_0}, E_{k_0}+e^{-4q_{n+1}\varepsilon_n^{\frac{1}{8}}})$, which
means that the gap with label $k_0$ is open. 
This finishes   the  proof of Theorem \ref{dry}.\qed \\


\section{Periodic approximation}

Here we discuss a variation of the previous approach based on
periodic approximation.  We will show
subexponential estimates for some of the gaps, and then use a simple
analysis of the projective action of the cocycles and of the
fibered rotation number. 

\subsection{Preliminaries}

\subsubsection{Variation of the most expanded direction}

Let $\bM(2,\R)$ be the space of two by two real matrices.
Let $\Omega \subset \bM(2,\R)$ be the set of $W$
with $|\det W|<\|W\|^2$.  For $W \in \Omega$, we
let $\nu_W$ be the eigenspace associated to the
largest eigenvalue (i.e. $\|W\|^2$) of $W W^*$.  Clearly $\nu_W$ is an
analytic function of $W \in \Omega$.

One can easily check that
\begin{Lemma}\label{variation}
There exists   an absolute constant $C_\Omega$ such that 
\begin{enumerate}
\item  The
derivative of $\nu_W$ is bounded by
$C_\Omega \frac {\|W\|} {\|W\|^2-|\det W|}$.
\item $A \in \SL(2,\R)$ and $W \in \Omega$ are such that
$\eta=\|A\|^2 \frac {|\det W|} {\|W\|^2}<1$, then
$W A \in \Omega$ and $\nu_{W A}$ and $\nu_W$ are $C_\Omega \eta$-close.
\end{enumerate}
\end{Lemma}

 As the proof is just direct computation, we leave the proof to the readers.  Also, we have the following:


\begin{Lemma} \label {W}

For every $\epsilon>0$ there exists $C_0$
with the following property.
Let $R>2$ and $W \in C^\omega_\epsilon(\R/\Z,\bM(2,\R))$ be such that
$\|W(z)\| \geq R^{-1}$ and $|\det W(z)| \leq R^{-3}$ whenever
$|\Im z|<\epsilon$.  Then the topological degree $l$
of $x \mapsto \nu_{W(x)}$ (as a function $\R/\Z \to \P \R^2$) satisfies
$|l| \leq C_0 \ln R.$

\end{Lemma}

\begin{pf}

Choose $C_1$ large.
Let $\tilde W$ be obtained by truncation of
the Fourier series of $W$ at level $C_1 \ln R$.  Then the eigendirections of
$\tilde W \tilde W^*$ are close to the ones of
$W^* W$ and $W W^*$, so the degrees of the eigendirections remain the
same.  If the horizontal direction is always an eigendirection, then the
degree is zero.  Otherwise the absolute value of the degree is bounded by
the number of times that the horizontal direction is an eigendirection,
which is the number of zeros of the (non-vanishing) lower left
coefficient and bounded in terms of the degree of $\tilde W$.
\end{pf}

\subsubsection{Lagrange interpolation and Convexity}

\begin{Theorem}

Let $\phi:\C/\Z \to \C$ be a trigonometric polynomial
\be
\phi(z)=\sum_{k_-
\leq k \leq k_+} \hat \phi_k e^{2 \pi i k z}.
\ee
Let $q>k_+-k_-$ and
$z_* \in \C/\Z$.  Then
\be
\sup_{\Im z=\Im z_*} |\phi(z)|^2 \leq \sum_{j=0}^{q-1} |\phi(z_*+j/q)|^2.
\ee

\end{Theorem}

\begin{pf}

Let $\Phi(x)=\sum_{j=0}^{q-1}|\phi(x+j/q+i \Im z_*)|^2$.
It is $1/q$-periodic and has only non-vanishing
Fourier coefficients of degree $|k| \leq k_+-k_-<q$, so it is constant and
$|\phi(z)|^2 \leq \Phi(z-i \Im z_*)=\Phi(z_*-i\Im z_*)$ when $\Im z=\Im
z_*$.
\end{pf}


By a Lagrange interpolation argument, we mean applying the above estimate to
bound an analytic function $\phi$ over a circle $\Im z=\Im z_*$ in terms of
the values over $z_*+j/q$, $0 \leq j \leq q-1$.  As a preliminary step one
must truncate the Fourier series of $\phi$ at an appropriate level
(determined by the size of $\phi$ over a neighborhood of the circle).

Given a subharmonic function $\phi$ (typically $\phi=\ln \|\Phi\|$ for a
vector of matrix valued holomorphic function) defined on a
band $t_-<|\Im z|<t_+$ in $\C/\Z$, Hadamard three-line theorem shows that  the function $t \mapsto
\sup_{\Im z=t} \phi(z)$ is convex over $t_-<t<t_+$.

\subsubsection{Projective action, Fibered rotation number}\label{fro}

Let us identify $\R/\Z$ with $\P\R^2$ by $y \mapsto \begin{pmatrix} \cos \pi
y\\ \sin \pi y\end{pmatrix}$.  Given
$A \in \PSL(2,\R)$ we denote by $y \mapsto A \cdot y$
the projective action $\R/\Z \to \R/\Z$.  Then we can write $A \cdot
y=y+\phi(y)$ for some analytic function $\phi\equiv\phi[A]:\R/\Z \to \R$.
Note that
$\phi$ is not uniquely defined, but two different choices differ by an
integer.  Moreover we have
\be \label {phi}
\max_{y \in \R/\Z} \phi(y)-\min_{y \in \R/\Z} \phi(y)<1.
\ee
It is impossible in general to make a consistent choice of $\phi$ over
all $A \in \PSL(2,\R)$ without discontinuities.    However this is possible in
one special case: if $A$ is of ``Schr\"odinger type'' in the sense that
$A \cdot 1/2=0$ we can select $\phi$ satisfying 
$\phi(1/2)=-1/2$.

Given $A \in C(\R/\Z,\PSL(2,\R))$ we can make a choice of $\phi_x \equiv
\phi[A(x)]$ depending continuously on $x \in \R$, which satisfy
$\phi_{x+1}=\phi_x+\deg A$.
So $\deg A=0$ when $A$ is homotopic to a constant,
and in this case $\phi_x$
can be defined (non-uniquely) over $x \in \R/\Z$.  Given such a choice, we
can make a choice $\phi_{x,k}$ for $\phi[A_k(x)]$ by letting
$\phi_{x,1}=\phi_x$ and imposing the cocycle rule
$\phi_{x,k+l}(y)=\phi_{x,k}(y)+\phi_{x+k \alpha,l}(A_k(x) \cdot y)$.
For positive $k$ this gives $\phi_{x,k}(y)=\sum_{m=0}^{k-1} \phi_{x+m
\alpha}(A_m(x) \cdot y)$.

The rotation number is defined as $ \tilde \rho \equiv  \tilde \rho[\alpha,A]$ by
$$  \tilde \rho_x=\lim_{k \to \infty} \frac {1} {k} \phi_{x,k}(y).$$  By (\ref {phi})
this is a uniform limit, so it
depends continuously on $x$, $\alpha$ and $A$.
It is however only
defined up to an integer, so it should be seen as a function $  \tilde \rho:\R/\Z \to
\R/\Z$.  Since $ \tilde \rho_x= \tilde \rho_{x+\alpha}$, it is actually a constant for
$\alpha \in \R \setminus \Q$, and we denote it by $\tilde \rho$.

Schr\"odinger cocycle are always
homotopic to a constant.   Indeed as $A^{(E)} \cdot 1/2=0$, which allows us to
make a consistent choice for $\phi[A^{(E)}(x)]$ by taking
$\phi[A^{(E)}(x)](1/2)=-1/2$.  Below we write $\phi^{(E)}_x$,
$\phi^{(E)}_{x,k}$, $ \tilde \rho_x(E)$ for the objects defined using this
choice.  

For every $y \in \R/\Z \setminus \{1/2\}$,
$E \mapsto A^{(E)}(x) \cdot y$ is an orientation reversing
diffeomorphism onto $\R/\Z \setminus \{0\}$.  So if $y \in (-1/2,1/2)$ then
$E \mapsto \phi^{(E)}_{x,k}(y)$ is an orientation reversing diffeomorphism
onto $(-k-y,-y)$, while $E \mapsto \phi^{(E)}_{x,k}(1/2)$ is an orientation
reversing diffeomorphism over $(-k+1/2,-1/2)$ for $k \geq 2$.
We conclude that for each $x \in \R/\Z$, $ \tilde \rho_x(E)$ is a non-decreasing
function taking values in $[-1,0]$.  In fact it is easy to see that for
$E>2+\sup_x v(x)$ the cocycle $(\alpha,A^{(E)})$
is uniformly hyperbolic and $   \tilde \rho (E)=-1$ while for $E<-2+\inf_x v(x)$ the
cocycle $(\alpha,A^{(E)})$ is uniformly hyperbolic and $  \tilde\rho {(E)}=0$.


 The integrated density of
states $N_{v,\alpha,x}$
gives the asymptotic distribution of eigenvalues of the restriction $H^I$
of $H$ to large intervals $I \subset \Z$.  More precisely, $N(E)$ is defined
as $\lim_{\#I \to \infty}
\frac {1} {\# I} \#\{j,\, E^I_j \leq E\}$, where $E^I_j$
are the eigenvalues of $H^I$.  Considering the cocycles $(\alpha,A^{(E)})$
associated to $v$ as before, we see that if $I=[a,b]$,
$E$ is an eigenvalue of $H^I$ precisely when $A^{(E)}_{b-a+1}(x+a \alpha)$
takes the horizontal direction to the vertical direction, that is, when
$\phi^{(E)}_{b-a+1}(0)=1/2 \mod \Z$.  The fact that the i.d.s. is well
defined is then just a consequence of the above analysis of the fibered
rotation number, which indeed establishes that 
\begin{equation*}
N(E)=-  \tilde  \rho(E).
\end{equation*}

Moreover, 
if $(\alpha,A)$ is uniformly hyperbolic we can compute the rotation number
using the unstable direction by taking some $B$ with $B(x) \cdot u(x)=0$
to get $\tilde \rho= \deg B \cdot  \alpha $ and observing that $-\deg B$ coincides with the
topological degree of $u$.  In particular $N(E)=k \alpha \mod \Z$,
which is the
statement of the Gap Labelling Theorem in
this setting.  We thus see that the label $k$ of the gap is just minus the
topological degree of the unstable (or stable) direction.

\subsubsection{Basis of Periodic Schr\"odinger  operators}

Let $q \geq 2$.
Given $v:\Z/q\Z \to \R$ we can consider the periodic Schr\"odinger operator
$$(H u)_n=u_{n+1}+u_{n-1}+v(n) u_n.$$  Let $A^{(E)}(n)=\begin{pmatrix}
E-v(n)&-1\\1&0 \end{pmatrix}$, $A^{(E)}_k(n)=A^{(E)}(n+k-1) \cdots
A^{(E)}(n)$.  Then $A^{(E)}_q(n)$ is conjugated to $ A^{(E)}_q(n+1)$ by
$A^{(E)}(n)$.

Let $\phi^{(E)}_n=\phi[A^{(E)}(n)]$ with $\phi^{(E)}_n(1/2)=-1/2$.
Let us note that
\be \label {d omega}
\frac {d} {dE} \phi^{(E)}_n(y)=\omega(   A^{(E)} (n)  \cdot y),
\ee
for some explicit analytic function $\omega:\R/\Z \to \R$ satisfying
$\omega( t) \leq 0$ with equality only at $ t=0$. Indeed, direct computation shows that $\omega (t)= -\frac{1}{\pi} \sin^2 \pi t  $.
 
Let $\phi_{n,k}^{(E)}(y)=\phi[A^{(E)}_k(n)]$ chosen so that
\be \label {phik}
\phi_{n,k}^{(E)}(y)=
\sum_{m=0}^{k-1} \phi^{(E)}_{n+m}(A^{(E)}_m(n) \cdot y).
\ee

For every $y \in \R/\Z \setminus \{1/2\}$,
$E \mapsto A^{(E)}(n) \cdot y$ is an orientation reversing
diffeomorphism onto $\R/\Z \setminus \{0\}$.  So if $y \in (-1/2,1/2)$ then
$E \mapsto  \phi^{(E)}_{n,k}(y) $ is an orientation reversing diffeomorphism
onto $(-k-y,-y)$, while $E \mapsto  \phi^{(E)}_{n,k}(1/2) $ is an orientation
reversing diffeomorphism over $(-k+1/2,-1/2)$ for $k \geq 2$.

It follows that for any integer
$0 \leq j \leq q$ there is an interval $G^j$ such
that $-j$ belongs to the image of $\phi^{(E)}_{0,q}$
and the $G^j$ are disjoint and ordered
from left to right in $\R$, while in the complement of $\bigcup_{j=0}^{q-1}
G^j$ the image of $\phi_{0,q}^{(E)}$ contains no integer.  The intervals
$G^0$ and $G^q$ are unbounded while the $G^j$, $1\leq j \leq q-1$ are
compact and called the $j$-th
gaps of $H$.  We call a gap collapsed if $G^j$ reduces
to a single energy $E$, and in this case $\phi^{(E)}_{0,q}=-j$ identically.

The set $\bigcup_{j=0}^q \inter G^j$ consists of the set of $E$ such that
$y \mapsto A^{(E)}_q(0) \cdot y$ has exactly two fixed points, so that
$A_q^{(E)}(0)$ is hyperbolic.  Its complement is the spectrum
$\sigma(H)$.

The complement of $\bigcup_{j=0}^q G^j$ is a union of $q$ open intervals
$\Delta^j$ such that the image of $\phi^{(E)}_{0,q}$
is contained in $(-j,-j+1)$. 
We call $\overline \Delta^j$ the $j$-th band of the spectrum.

Let $a(E)=\tr A^{(E)}_q(0)$.  For $E \in \Delta^j$,
$y \mapsto A^{(E)}_q(0) \cdot y$ has no fixed point so $|a(E)|<2$, and
$\frac {d} {dE} A^{(E)}_q(0) \cdot y<0$ then implies that
$\frac {d} {dE} a(E) \neq 0$.  Since at the boundary of
$\Delta^j$, $y \mapsto A^{(E)}_q(0) \cdot y$ has fixed
points, we conclude that $a$ restricts to a diffeomorphism from $\Delta^j$
to $(-2,2)$.  Note that $a(E)$ is a monic polynomial of degree $q$, so we
conclude that each $G^j$, $1 \leq j \leq q-1$,
contains exactly one critical point of $a(E)$ (necessarily non-degenerate). 

For $1 \leq j \leq q-1$ and $E \in \partial G^j$,
$\tr A^{(E)}_q(0)=(-1)^{q-j} 2$ on $\partial
G^j$, and, as discussed before, $G^j$ is collapsed if
$A^{(E)}_q(0)=(-1)^{q-j} \id$.  In the non-collapsed case we have the
following quantitative
estimate:

\begin{Lemma} \label {gaps}

Let $1 \leq j \leq q-1$ and let $E_j \in \partial G^j$.  Then
\begin{equation}
|G^j| \geq C_1 \min \{\|A^{(E_j)}_q(0)-(-1)^{q-j} \id\|,1\} g^{-1}_j
\end{equation}
for some absolute constant $C_1$ and with
\begin{equation}
g_j=\sup_{E \in G^j} \|\sum_{k=0}^{q-1} (A^{(E)}_k(-k)^{-1})^*
A^{(E)}_k(-k)^{-1}\| \leq \sup_{E \in G^j}
\sum_{k=0}^{q-1} \|A^{(E)}_k(-k)\|^2.
\end{equation}

\end{Lemma}

\begin{pf}

Assume for definiteness that $E_j$ is the left boundary of $G^j$.  Then
$ \phi^{(E_j)}_{0,q}(\R/\Z)$ is an interval $[-j,-j+\epsilon_j]$ where
\be
C_2^{-1} \leq
\frac {\epsilon_j} {\min \{\|A^{(E_j)}_q(0)-(-1)^{q-j} \id\|,1\}} \leq C_2
\ee
for some absolute constant $C_2$. To prove this, we can just assume $$A^{(E_j)}_q(0)= (-1)^{q-j} \begin{pmatrix} 1& c \\0&1 
\end{pmatrix},$$ which gives $ A^{(E_j)}_q(0) \cdot 0 =  \phi^{(E_j)}_{0,q}(0) = - j$, direct computation shows that  $\epsilon_j$ is of size $|c|$.

On the other hand at the
right boundary $E_j'$ we have
$\max_{y \in \R/\Z}   \phi^{(E'_j)}_{0,q}(y)=-j$.  Thus it remains to show
that $  \max_{y \in \R/\Z}  |\frac {d} {dE} \phi^{(E)}_{0,q}(y)| \leq C_3 g_j$ for some absolute
constant $C_3$.

Let $y_0 \in \R/\Z$, and for $1 \leq k \leq q$ let $\gamma_k(y)=\partial_y
A^{(E)}_{q-k}(k) \cdot y$ and $y_k=A^{(E)}_k(0) \cdot y_0$.  Then we have
\be
\partial_E \phi^{(E)}_{0,q}(y_0)=\sum_{k=1}^q \omega(y_k)
\gamma_k(y_k).
\ee
Indeed,  by (\ref {d omega}) and \eqref{phik},   we have 
\begin{eqnarray*}
\partial_E \phi^{(E)}_{0,k}(y_0) &=& \partial_E \phi^{(E)}_{0,k-1}(y_0)+  \partial_E \phi^{(E)}_{k-1}( y_{k-1}) + (\partial_y A^{(E)}(k-1) \cdot y_{k-1}  -1  ) \partial_E \phi^{(E)}_{0,k-1}(y_0) \\
&=& \omega (y_k)+ \partial_y A^{(E)}(k-1) \cdot y_{k-1}  \partial_E \phi^{(E)}_{0,k-1}(y_0) ,
\end{eqnarray*}
then the result follows by iteration.

On the other hand, direct computation shows that  for any $A\in \SL(2,\R)$, $ \partial_y A\cdot y =\frac{1}{\|Ay\|^2}$. Thus if $w$ is a unit vector in the direction of $y_q$, then
$\gamma_k(y_k)=\|A_{q-k}(k)^{-1}
w\|^2$, which gives the desired bound with $C_3=-\min_{y \in \R/\Z}
\omega(y)$.

\end{pf}

\subsection{Periodic almost Mathieu operators}

We recall now elements of the theory of periodic almost Mathieu operators. 
The basic reference is \cite {AvMS}.
Given $\lambda>0$, $p/q \in \Q$ and $x \in \R/\Z$, let
$$(H_{\lambda,p/q,x} u)_n=u_{n+1}+u_{n-1}+2 \lambda \cos 2 \pi (x+n
p/q) u_n.$$  Those are periodic operators and we denote
$\Sigma_{\lambda,p/q,x}=\sigma(H_{\lambda,p/q,x})$.  We are also interested
in the joint spectrum
$\Sigma_{\lambda,p/q}=\bigcup_{x \in \R/\Z} \Sigma_{\lambda,p/q,x}$.

Let us denote by $a_{(\lambda,p/q,x)}(E)$, $G^j_{\lambda,p/q,x}$ and
$\overline \Delta^j_{\lambda,p/q,x}$ the objects associated to each
$H_{\lambda,p/q,x}$ as in the previous section.

The function $x \mapsto a_{\lambda,p/q,x}(E)$ is a trigonometric polynomial
with vanishing Fourier coefficients of degree $k$ for $|k|>q$. 
Its Fourier coefficients of degree $\pm q$ are
readily computed to be $\prod_{j=0}^{q-1} -\lambda e^{\pm 2 \pi i j p/q}=
-\lambda^q$.  Moreover
it is $1/q$-periodic, so we must have $a_{(\lambda,p/q,x)}(E)=-
2 \lambda^q \cos 2 \pi q x+a_{\lambda,p/q,1/4q}(E)$.  This is known as
Chambers formula.

The Chambers formula implies that the different
$G^j_{\lambda,p/q,x}$
are nested, with the largest and the smallest one
corresponding to critical points $x$ of $\cos 2 \pi x$. 
Specifically, the smallest $G^j_{\lambda,p/q,x}$ corresponds to maxima of
$(-1)^{q-j} \cos 2 \pi x$.  Writing $x_{j,q}=0$ if $q-j$ is even and
$x_{j,q}=1/2q$ if $q-j$ is odd, we then have
$G^j_{\lambda,p/q}=\bigcap_{x \in \R/\Z}
G^j_{\lambda,p/q,x}=G^j_{\lambda,p/q,x_{j,q}}$, $0 \leq j \leq q$.
On the other hand,
$\Delta^j_{\lambda,p/q}=\bigcup_{x \in \R/\Z} \Delta^j_{\lambda,p/q,x}$, $1
\leq j \leq q$ is
the space between $G^{j-1}_{\lambda,p/q}$ and $G^j_{\lambda,p/q}$ and are
hence disjoint.

The $\overline \Delta^j_{\lambda,p/q}$, $1 \leq j \leq q$, are called the
$j$-th bands
of $\Sigma_{\lambda,p/q}$ while $G^j_{\lambda,p/q}$, $1 \leq j \leq q-1$ are
called the $j$-th gaps of $\Sigma_{\lambda,p/q}$.  The gaps are called
collapsed or non-collapsed as before.

While there are indeed many exponentially small
gaps, in the following, we show that there are
subexponential estimates for some of the gaps:

\begin{Theorem} \label {subex}

Let $\alpha \in \R \setminus \Q$ and $0<\lambda_0<\lambda_1<1$.  For every
$\delta>0$ there exists $\delta_*>0$ with the following property.  Let
$\lambda_0<\lambda<\lambda_1$, $p/q
\in \Q$, $1 \leq j \leq q-1$ and
$k \in \Z$ be such that $|\alpha-p/q|<\delta_*$,
$|k|<\delta_* q$ and $j=k p \mod q$.  Then $|G^j_{\lambda,p/q}| \geq
e^{-\delta q}$.

\end{Theorem}

In the rational case $\alpha=p/q$, for each $x$ one obtains a periodic
Schr\"odinger operator.  In this case we see that $N$
maps the $j$-th band homeomorphically onto
$[(j-1)/q,j/q]$, while being constant equal to $j/q$ on the
$j$-th gap.  It is then natural to consider the label of the gap as an
integer $k$ such that $j/q=k p/q$,
but this is only defined $\mod q$.  We thus see that Theorem \ref {subex}
concerns lower bounds on the size of gaps, which are subexponential on the
distance from the label $k$ to $q \Z$. \\

\smallskip
\textbf{Proof of Theorem \ref{dry} in the $\beta>0$ case:} 
By duality, it suffices to consider
$H_{\lambda,\alpha,x}$ for $0<\lambda<1$.  Then the result follows directly
from $\frac{1}{2}$-H\"older continuity of the
gaps (Theorem 7.3 of \cite{AvMS}) and Theorem \ref{subex}.\qed\\

\subsection{Proof of Theorem \ref {subex}}

The proof of Theorem \ref {subex} depends on Quantitative Almost
Reducibility, expressed in Theorem \ref {exp}, and Quantitative Aubry
Duality.  Let us first establish each of those.

\subsubsection{Quantitative Almost Reducibility: Proof of Theorem \ref {exp}}\label{exp-proof}

We will use notation $o(\cdot)$ for quantities that become small compared
with $\cdot$ provided that $q_*$ is large and $\delta_*$ is small.
For instance, (\ref {sizek}) can be written as
$\sup_{0 \leq k \leq q} \|A_k(0)\|_{\epsilon_0} \leq e^{o(1) q}$.  Iterating gives
$A_k(z)A_q(z)=A_q(z+k\frac{p}{q})A_q(z)$, consequently  
\be
\sup_{0 \leq k \leq q-1}
\|A_q(k p/q)-I\| \leq e^{(-c_0+o(1)) q}.
\ee
A Lagrange interpolation argument then implies that
\be
\|A_q(z)-I\|_0 \leq e^{(-c_0+o(1)) q}.
\ee
Using (\ref {sizek}) again and convexity gives
\be
\|A_q(z)-I\|_\epsilon \leq e^{(-\gamma c_0+o(1)) q}.
\ee
This kind of argument will play a role several times below, and will be
referred as the Lagrange interpolation and convexity argument.

If $I=\id$ let $\bp=0$, otherwise let $\bp$ be defined as follows.  Let $0
\leq \bq \leq q-1$ be such that $\bq p=1 \mod q$, and let
$\bp=1-2 [\bq/2] p$.  Then $\bq \bp=1 \mod 2q$ if $\bq$ is odd and $\bq
\bp=0 \mod 2q$ if $\bq$ is even.  Set $\bw=0$ if $I=\id$ or $\bq$ is even
and $\bw=1$ otherwise, so that $\bp \bq=\bw \mod 2q$.
Consider the ``discrete Fourier
transform'' of the sequence $\{R_{k \bp/2q}
A_k\}_{k=0}^{q-1}$ given by
$W_k=\sum_{s=0}^{q-1} R_{ksp/q} R_{s \bp/2q} A_s$,
$0 \leq k \leq q-1$.  Then
\be \label {conj}
W_k(z+p/q) A(z)=R_{-k p/q} R_{-\bp/2q} (W_k(z)+I A_q(z)-\id).
\ee

One easily checks Fourier inversion
\be
R_{k \bp/2q} A_k=\frac {1} {q}
\sum_{s=0}^{q-1} R_{-ksp/q} W_s, \quad 0 \leq k \leq q-1.
\ee
Let $J$ be the set of all $0 \leq k \leq q-1$ such that $\|W_k(0)\| \geq
\frac {1} {q}$.  By Fourier inversion $\sum_{k=0}^{q-1} W_k(z)=q \id$.  It
follows that $\|\frac {1} {q} \sum_{k \in J} W_k(0)-\id\|
\leq \frac {1} {q}$.  In particular $J$ is non-empty.


Note that if $0<\delta_0<\gamma c_0$,
an exponential bound $\|W_k(z_*)\| \leq e^{(-\delta_0+o(1)) q}$
for some $z_*$ with
$|\Im z_*|<\epsilon$ gives an exponential bound $\|W_k(z)\|_\epsilon
\leq e^{(-\gamma \delta_0+o(1)) q}$.
Indeed (\ref {conj}) implies that $\|W_k(z_*+j p/q)\| \leq
e^{-(\delta_0+o(1)) q}$ for $0 \leq j
\leq q-1$, so we can conclude by
Lagrange interpolation and convexity.  In particular, $\|W_k(z)\| \geq
e^{-o(q)}$ for $k \in J$.

Similarly,
if we denote $w_k=\det W_k$,
\eqref{conj} gives
\be \label{error-det}
\|w_k(z+\frac{p}{q})-w_k(z)\|_\epsilon \leq e^{(-\gamma
c_0+o(1)) q},
\ee
thus   if $\|w_k(z_*)\| \leq e^{(-\delta_0+o(1))q}$
for some $z_*$ with
$|\Im z_*|<\epsilon$ then $\|w_k(z)\|_\epsilon \leq e^{(-\gamma
\delta_0+o(1)) q}$.

For $k \in J$, let
$R_k=\max \{q^2
\|A\|^2_0,\sup_{|\Im z|<\epsilon} \|W_k(z)\|^{-1}\}$. 
In particular $R_k \leq e^{o(q)}$.
Assume that
$\|w_k\|_\epsilon \geq R_k^{-3}$, then above discussion implies that $\inf_{|\Im z|<\epsilon} |w_k(z)|
\geq e^{-o(q)}$.
If $w_k(0)>0$, let $b(z)^2=w_k(z)$ and let
$B(z)=\frac {1} {b(z)} W_k(z)$, and if $w_k(0)<0$, let
$b(z)^2=-\det W_k(z)$ and let $$B(z)=\frac {1} {b(z)}
\begin{pmatrix} 1&0\\0&-1
\end{pmatrix} W_k(z).$$  Then $\|B\|_\epsilon \leq e^{o(q)}$, and (\ref{error-det}) implies that 
$B(z+p/q) A(z) B(z)^{-1}$ is $e^{(-\gamma c_0+o(1)) q}$
close   to $R_{-k p/q}$    on $|\Im z|<\epsilon$, and it is
$e^{(-c_0+o(1)) q}$
close on $\Im z=0$, which gives the result.

Assume now that
$\|w_k\| \leq R_k^{-3}$ for $k \in J$.
Since $\det (\frac {1} {q} \sum_{k \in J} W_k(0)) \geq 1-o(1)$,
$J$ must have at least two elements.




Note that by Lemma \ref{variation},  (\ref {conj}) gives that
$\nu_{W_k(x+p/q)}$ and
$R_{-k p/q} R_{-\bp/2q} \nu_{W_k(x)}$ are $2 C_\Omega q^{-2}$ close.
By iteration $\bq$ times, Lemma \ref{variation} gives that $\nu_{W_k}(x+1/q)$ and
$R_{-k/q} R_{-\bw/2q}$ $\nu_{W_k}(x)$ are $2 C_\Omega q^{-1}$ close.

The choice of $\bp$ was made in order to have
$R_{\bw/2q}$ is  $o(1)$ close to $\id$, which plays a role in the following
argument.
Let $R_{\phi_k(x)} \begin{pmatrix} 1\\0 \end{pmatrix} \in
\nu_{W_k}(x)$ with $\phi_k:\R \to \R$ continuous.  Let $l_k$ be the
topological degree of $x \mapsto \nu_{W_k(x)}$.  Then
$\phi_k(x+1)=\phi_k(x)+\frac {1} {2} l_k$ and
$2(\phi_k(x+1/q)-\phi_k(x)+\frac {k} {q}+\frac {\bw} {2q})$
is $4 C_\Omega q^{-1}$ close to an integer.
By Lemma \ref {W}, $|l_k|=o(q)$, so
$-2 k/q$ is $o(1)$ close to an integer.  Since
$0 \leq k \leq q-1$, this means that $k/q$ must be
close to $r_k$ with $r_k \in \{0,1/2,1\}$.

Let $J_1 \subset J$ be the set of $k$ with $r_k$ either $0$ or $1$, and let
$J_2 \subset J$ be the set of $k$ with $r_k=1/2$.





Let $M_k(x)=R_{(k-q r_k) x} W_k(x)$, $x \in \R$ so that
$M_k(x+1)=(-1)^{2 q r_k} M_k(x)$.
We have that $M_k(x+p/q) A(x)$ is
exponentially close to $(-1)^{2 p r_k} R_{-\bp/2q} M_k(x)$.

Let $L_j=\frac {1} {q} \sum_{j \in J_j} M_j$, $j=1,2$.  Let $L^+=L_1+L_2$ and
$L^-=L_1-L_2$.
We have
$L_1(z+1)=L_1(z)$ and $L_1(z+p/q) A(z)$ is exponentially close to
$R_{-\bp/2q} L_1(z)$
in $|\Im z|<\epsilon$.  On the other hand, $L_2(z+1)=(-1)^q
L_2(z)$ and $L_2(z+p/q) A(z)$ is exponentially close to $(-1)^p R_{-\bp/2q}
L_2(z)$.  Also $L^+(0)=\frac {1} {q}
\sum_{k \in J} W_k(0)$ is $1/q$ close to $\id$, so
$\det L^+(0)$ is close to $1$.

Note that $L_1$, $L_2$, $L^+$ and $L^-$ are $2$-periodic, and their
determinant along $2 k p/q$, $0 \leq k \leq q-1$, is exponentially close to
their determinant at $0$.  By Lagrange interpolation and convexity, the same
holds throughout $|\Im z|<\epsilon$.

If $q$ is odd then $L^-(z+1)=L^+(z)$,  and
if $q$ is even (then $p$ is odd), $L^-(z+p/q) A(z)$ is exponentially close to $R_{-\bp/2q}
L^+(z)$.  So $\det L^+(0)$ is close to $\det L^-(0)$.

Note that $\det L^+ + \det L^-=2(\det L_1+\det L_2)$.  Since $\det L^+(0)$
is close to $1$, this implies that $\det L_1(0)+\det L_2(0)$ is close to
$1$.  In particular, for either $j=1$ or $j=2$ we have $\det L_j(0) \geq
1/3$.

Let $b(z)^2=\det L_j$ and take $B(z)=\frac {1} {b(z)}
L_j(z)$, if $j=1$ or $q$ is odd, and $B(z)=\frac {1} {b(z)} R_{-z/2} L_j(z)$
otherwise (the extra rotation is introduced to bring the conjugacy to
$\SL(2,\R)$).

For the last statement, let $l$ be such that $B$ is homotopic to $x \mapsto
R_{l x}$.  Then $l$ can be estimated by counting (after truncation of the
Fourier series) zeros of coefficients of
$B$, similarly to Lemma \ref {W}, and we get $|l| \leq o(q)$.
We can then replace $B$ by $x \mapsto R_{-lx} B(x)$
to obtain a conjugacy homotopic to a
constant. 
\qed

\subsubsection{Quantitative  Aubry duality argument}
In this subsection, we develop a  \textit{Quantitative version of Aubry duality} for rational frequency, which is parallel to the ones in Section \ref{qad}. 
Recall that
\be 
A^{(\lambda,E)}(x)=\begin{pmatrix} E-2 \lambda \cos 2 \pi x & -1
\\ 1 & 0 \end{pmatrix}.
\ee

\begin{Lemma} \label {exp3}

For every $\epsilon>0$, $c>0$, $\lambda_0>0$, there exists $\delta_*>0$ and
$q_*>0$ with the following property.  Let $q>q_*$, $\lambda_0<\lambda<1$ and
$E \in [-5,5]$, $I \in \{\id,-\id\}$ and $A=A^{(\lambda,E)}$.
Then there is no $B \in C^\omega_\epsilon(\R/\Z,\PSL(2,\R))$ with
\be \label {ests}
\|B(x+p/q)A(x)B(x)^{-1}-
I\|_\epsilon \leq e^{(-c+\delta_*) q} \quad \text
{and} \quad \|B\|_\epsilon \leq e^{\delta_* q}.
\ee

\end{Lemma}

\begin{pf}

We use the notation $o(\cdot)$ as before.  Let $\varepsilon \in \{-1,1\}$ be
such that $I=\varepsilon \id$.  Arguing by contradiction, assume
that there exists $B$ satisfying (\ref {ests}).
Then $\|A(x) B(x)^{-1}-\varepsilon B(x+p/q)^{-1}\|_\epsilon
\leq e^{(-c+o(1)) q}$.  If $B(x)^{-1}=\begin{pmatrix}
u^+&u^-\\ v^+&v^- \end{pmatrix}$ this gives
$\|v^\pm(x+p/q)-\varepsilon u^\pm(x)\|_\epsilon \leq
e^{(-c+o(1))q}$ and $$(E-2 \lambda \cos 2 \pi x)
u^\pm(x)-\varepsilon  u^\pm(x+p/q)-\varepsilon  u^\pm(x-p/q)\|_\epsilon
\leq e^{(-c+o(1)) q}.$$
Let $\m$ be either $0$ or $1$ according to whether $B$ takes values in
$\SL(2,\R)$ or not.  Taking the Fourier series
$u^\pm(x)=\sum_{k \in \Z} \hat u^\pm_k e^{(2 k+\m) \pi i x}$
we get
$$|(E-2 \varepsilon \cos (2 k+\m) \pi p/q) \hat u^\pm_k-
\lambda (\hat u^\pm_{k+1}+\hat
u^\pm_{k-1})| \leq e^{(-c+o(1)) q-2 \pi \epsilon |k|}.$$

Setting
$\A(x)=A^{(1/\lambda,\varepsilon E/\lambda)}$,
$\U^\pm_k=\begin{pmatrix} \varepsilon^k \hat u^\pm_k
\\ \varepsilon^{k-1} \hat u^\pm_{k-1} \end{pmatrix}$,
and $\U_k$ the matrix with columns $\U^+_k$ and
$\U^-_k$, we have
\be \label {dec}
\|\U_k\| \leq e^{o(q)-2 \pi \epsilon k}
\ee
and
\be \label {jump}
\|\A(k p/q+\m p/2q) \U_k-\U_{k+1}\| \leq e^{(-c+o(1)) q-2 \pi \epsilon |k|}.
\ee

Similar as Lemma \ref{z1-estimate}, we have $\sum_{k \in \Z} \|\U^\pm_k\|^2 =2 \int_{\R/\Z} |u^\pm(x)|^2 dx
\geq e^{-o(q)}$.  Thus (\ref {dec}) gives that there exists $k_\pm$ such
that $|k_\pm| \leq o(q)$ with $\|\U^\pm_{k_\pm}\| \geq e^{-o(q)}$.
Letting $C_0$ be such that $\ln \|\A\|_0 \leq C_0$ (where we use $\lambda>\lambda_0$),
(\ref {jump}) gives, for any small $\delta_0>0$,
\be \label {U}
\|\U^\pm_k\| \geq e^{(-C_0 \delta_0-o(1)) q}, \quad |k| \leq
(\delta_0-o(1)) q.
\ee

Using (\ref {dec}) 
and (\ref {jump}), we get
\be \label {jump1}
|\det \U_k-\det \U_{k+1}| \leq e^{(-c+o(1)) q-4 \pi \epsilon |k|}.
\ee
Using  (\ref {jump1}) we get
\be \label {U1}
|\det \U_k| \leq e^{(-c+o(1)) q}.
\ee
Now (\ref {U}), (\ref {U1}) imply that for $|k| \leq (\delta_0-o(1)) q$
we can write
$\U^+_k=\gamma_k \U^-_k+\V_k$ with $e^{(-C_0 \delta_0-o(1)) q}
\leq |\gamma_k| \leq e^{(C_0 \delta_0+o(1)) q}$ and $\V_k$
orthogonal to $\U^-_k$ satisfying $\|\V_k\| \leq e^{(-c+C_0 \delta_0+o(1))
q}$.  This implies by (\ref {jump}) that $|\gamma_{k+1}-\gamma_k| \leq
e^{(-c+C_0 \delta_0+o(1)) q}$, for $|k| \leq (\delta_0-o(1)) q$ so that
\be
\|\U^+_k-\gamma_0 \U^-_k\| \leq e^{(-c+C_0 \delta_0+o(1)) q}, \quad |k| \leq
(\delta_0-o(1)) q.
\ee
Taking the Fourier transform we get $\|u^+-\gamma_0 u^-\|_0 \leq
e^{(-\delta_1+o(1)) q)}$ so that
$$|\det B(x)^{-1}| = \det \begin{pmatrix}
u^+ - \gamma_0 u^- &  u^-\\ v^+ - \gamma_0 v^- &  v^- \end{pmatrix}  \leq e^{(-\delta_1+o(1)) q},$$
providing the desired contradiction.
\end{pf}

\subsubsection{Proof of Theorem \ref {subex}}

Let $A^{(\lambda,E)}(x)$ be given by (\ref {alambdae}) and let
$A^{(\lambda,\alpha,E)}_k(x)=
A^{(\lambda,E)}(x+(k-1) \alpha) \cdots A^{(\lambda,E)}$.  We let
$\Sigma_{\lambda,\alpha,x}$ be the spectrum of the almost Mathieu operator
$(Hu)_n=u_{n+1}+u_{n-1}+2 \lambda \cos 2 \pi (x+n \alpha) u_n$ (which is
independent of $x$ for $\alpha \in \R \setminus \Q$, and
$\Sigma_{\lambda,\alpha}=\bigcup_{x \in \R/\Z} \Sigma_{\lambda,\alpha,x}$.

For fixed $\alpha \in \R \setminus \Q$ we have by
Theorem \ref {bj-formula} (see also
for instance Proposition 3.1 of \cite {AYZ}) the estimate
\be
\|A^{(\lambda,\alpha,E)}_k(x)\|_{-\ln \lambda}
\leq e^{o(k)}
\ee
uniformly over $\lambda_0<\lambda<1$ and $E$ in $\Sigma_{\lambda,\alpha}$.
Continuity of the spectrum and general uppersemicontinuity then gives
\be \label {subex1}
\max_{0 \leq k \leq q} \|A^{(\lambda,p/q,E)}_k\|_{-\ln \lambda} \leq e^{o(q)}
\ee
if $|\alpha-p/q|=o(1)$ and $E$ is $o(1)$ close to $\Sigma_{\lambda,p/q}$.

Let us consider the $j$-th gap $G^j_{\lambda,p/q}=[E_-,E_+]$ for some $j$
satisfying $j=k p \mod q$ with $|k|=o(q)$.
At $E_\pm$ we have $(-1)^{q-j} a_{\lambda,p/q,x_{j,q}}=2$,
while at $(E_-,E_+)$ we have $|(-1)^{q-j} a_{\lambda,p/q,x_{j,q}}|>2$.

Denote $A^{(E)}=A^{(\lambda,E)}$, $A^{(E)}_k=A^{(\lambda,p/q,E)}_k$ for short.
Let $\epsilon_0=-\ln \lambda$, let $0<\epsilon<-\ln \lambda$.  Let us show
that
\be \label {subex2}
\|A^{(E_\pm)}_q(x_{j,q})-(-1)^{q-j} \id\|
\geq e^{-o(q)}.
\ee
Fix some small $c>0$.  By (\ref {subex1}) and
Theorem \ref {exp}, if $\|A^{(E_\pm)}_q(x_{j,q})-(-1)^{q-j} \id\|
\leq e^{-c q}$ then there exists
$B' \in C^\omega_\epsilon(\R/\Z,\SL(2,\R))$ homotopic to a constant
such that $\|B'\|_\epsilon \leq e^{o(q)}$ and $\tilde
A(x)=B'(x+p/q)A^{(E_\pm)}(x)B'(x)^{-1}$ satisfies
$\|\tilde A-R\|_\epsilon \leq e^{(-\gamma c+o(1)) q}$ with some
$\gamma>0$ and $R^q=(-1)^{q-j} \id$.
By  the new definition of the fibered rotation number (c.f. Section \ref{fro}), the matrix $R$ must be in fact
$R_{(q+j)/2q}$.  We can replace $B'$ by $B(x)=R_{-k x/2} B(x)$ to obtain a
conjugacy close to $\pm \id$.  Since $|k| \leq o(q)$,
$\|R_{-kx/2}\|_\epsilon \leq e^{o(q)}$.  This contradicts Lemma
\ref {exp3} and proves (\ref
{subex2}).

Now if $|G^j_{\lambda,p/q}|=o(1)$ then for $E \in G^j_{\lambda,p/q}$ we have
(\ref {subex1}).  By Lemma \ref {gaps} and (\ref {subex2}), this implies
that $|G^j_{\lambda, p/q}| \geq e^{-o(q)}$ as desired.
\qed

\section*{Acknowledgements}
A.Avila was partially supported by the ERC Starting Grant \textquotedblleft Quasiperiodic\textquotedblright  and a grant from the SNSF. J.You and Q. Zhou were  partially supported by National Key R\&D Program of China (2020 YFA0713300) and Nankai Zhide Foundation.  J. You was also partially supported by NSFC grant (11871286). Q. Zhou was supported by NSFC grant (12071232).

\end{document}